\documentclass{llncs}
\usepackage{graphicx}
\usepackage{wrapfig}

\graphicspath{{./figs/}}
\usepackage{url}

\usepackage{extarrows}
\usepackage{graphics}
\usepackage{multirow}
\usepackage{amsfonts}
\usepackage{stmaryrd}
\usepackage{graphicx}
\usepackage{booktabs}
\usepackage{xparse}

\usepackage[ruled,linesnumbered,nofillcomment]{algorithm2e}

\usepackage{color}
\usepackage{subfigure}
\usepackage{amssymb}
\usepackage{hyperref}
\usepackage{algorithmic}
\usepackage{amsmath}
\usepackage[fleqn,tbtags]{mathtools}
\usepackage{txfonts}
\usepackage{float}

\usepackage{listings}

\usepackage{wasysym}
\usepackage{enumerate}

\makeatletter
\begingroup \catcode `|=0 \catcode `[= 1
\catcode`]=2 \catcode `\{=12 \catcode `\}=12
\catcode`\\=12 |gdef|@xcomment#1\end{comment}[|end[comment]]
|endgroup
\def\@comment{\let\do\@makeother \dospecials\catcode`\^^M=10\def\par{}}
\def\begincomment{\@comment\@xcomment}
\makeatother

\usepackage{url}
\usepackage{cite}

\newcounter{NumofComms}

\def\err{\mathit{Err}}

\newcommand{\set}[1]{\{#1\}}

\newcommand{\srct}[1]{\texttt{#1}}

\newcommand{\srci}[1]{$\mathit{#1}$}

\newsavebox{\mylistingbox}
\newsavebox{\mylistingboxx}

\newsavebox{\mylistingboxa}
\newsavebox{\mylistingboxb}
\newsavebox{\mylistingboxc}

\def\ao{\mathit{O}}

\newcommand{\roles}{\mathit{R}}

\def\preobj#1{[#1]}
\def\nd{nd}
\def\err{err}

\def\ncallee{\mathbf{callee}}
\def\narg#1{\mathbf{arg_{#1}}}

\def\nscope#1{\mathbf{scope_{#1}}}

\newcommand\oname[2]{\mathit{#1 [#2]}}
\newcommand\pname[1]{\mathit{#1}}
\newcommand\rname[1]{\mathit{#1}}

\newcommand\multmap[1]{\mathbf{#1}}
\def\packagej{\mathcal{P}}
\def\snapshot{\mathit{sp}}

\def\cname{\mathit{name}}
\def\atts{\mathit{Atts}}
\def\creators{\mathit{M_C}}
\def\modifiers{\mathit{M_M}}

\def\coid{\mathit{id}}
\def\coclass{\mathit{class}}

\def\co{\mathit{o}}
\def\rolename{\mathit{rname}}

\def\invoc{\mathit{invoc}}
\def\srcsp{\mathit{sp_s}}
\def\destsp{\mathit{sp_d}}
\def\roles{\mathit{Roles}}

\def\ao{\mathit{ao}}
\def\preds{\mathit{Preds}}

\def\equivalentao{\equiv}

\def\hg{\mathit{hg}}
\def\hgnodes{\mathit{V}}
\def\hgedges{\mathit{E}}
\def\hgsrc{\mathit{s}}
\def\hgdest{\mathit{t}}
\def\hgedgelabel{\mathit{l}}

\def\sp{\mathit{sp}}

\def\nao{\mathit{nao}}
\def\naoid{\mathit{id}}
\def\plural{\mathit{pl}}
\def\injective{\mathit{nj}}

\def\equivalentnao{\equiv}

\def\smallernao{\prec}

\def\smallereqnao {\preceq}

\def\srcnao{\mathit{{\nao}_1}}
\def\destnao{\mathit{{\nao}_2}}
\def\mult{\mathit{u}}

\def\smallereqmap {\preceq}

\def\ng{\mathit{ng}}
\def\ngnodes{\mathit{V}}
\def\ngedges{\mathit{E}}
\def\ngsrc{\mathit{s}}
\def\ngdest{\mathit{t}}
\def\ngedgelabel{\mathit{l}}

\def\smallereqng{\preceq}

\def\rng{\mathit{cng}}
\def\mrng{\mathit{cng_m}}
\def\rngnodes{\mathit{V}}
\def\rngedges{\mathit{E}}
\def\rngsrc{\mathit{s}}
\def\rngdest{\mathit{t}}
\def\rngedgelabel{\mathit{l}}
\def\rngrole{\mathit{u}}

\def\method{\mathit{m}}
\def\exceptionname{\mathit{e}}
\def\srcng{\mathit{\ng}}
\def\destng{\mathit{\ng'}}
\def\srcrng{\mathit{\rng}}
\def\destrng{\mathit{\rng'}}
\def\mapng{\mathit{p}}
\def\maprng{\mathit{q}}

\def\smallereqrule{\preceq}

\def\snapshots{\mathit{Sps}}

\def\calleeobj{\mathit{o_{callee}}}
\def\actualparams{\mathit{params}}
\def\raos{\mathit{raos}}
\def\naos{\mathit{naos}}
\def\block{\mathit{block}}
\def\partition{\mathit{partition}}
\def\coarsestpartition{\mathit{coarsestPartition}}
\def\coarsestpartitionm{\mathit{partition}}
\def\edgelabels{\mathit{L_E}}
\def\objectsorder{\mathit{orderedSet}}
\def\rules{\mathit{Rules}}
\def\mrule{\mathit{r_m}}

\newcommand{\subgraph}[2]{\mathit{subgraph(#1,#2)}}
\newcommand{\mergeable}[2]{\mathit{mergeable(#1,#2)}}
\newcommand{\roleconsistent}[2]{\mathit{roleconsistent(#1,#2)}}
\newcommand{\range}[1]{\mathit{range(#1)}}

\newcommand{\hgtong}[1]{\mathit{hton(#1)}}
\newcommand{\reachablefrom}[2]{\mathit{ReachableFrom(#1,#2)}}
\newcommand{\reachingto}[2]{\mathit{ReachingTo(#1,#2)}}
\newcommand{\reachablefromto}[2]{\mathit{ReachableToFrom(#1,#2)}}
\newcommand{\reachablespanning}[2]{\mathit{ReachingUndirected(#1,#2)}}
\newcommand{\renest}[2]{\mathit{renest(#1,#2)}}

\def\explorealg{\mathsf{Explore}}
\def\prunealg{\mathsf{PruneRedundants}}
\def\combinengs{\mathsf{Merge_N}}
\def\combinerngs{\mathsf{Merge_R}}
\def\combinemaps{\mathsf{CombineMappings}}
\def\mergeallalg{\mathsf{MergeAll}}

\def\mergealg{\mathsf{Merge}}
\def\lumpng{\mathsf{Lump}}
\def\lumpngm{\mathsf{LumpMergeds}}
\def\lumpngfinal{\mathsf{LumpFinal}}
\def\createrulealg{\mathsf{CreateRule}}

\def\completifyalg{\mathsf{CompletifyRules}}
\def\hgtongalg{\mathsf{TransfertoNested}}

\def\findsimilars{\mathsf{FindSimilars}}
\def\partitioncoarsest{\mathsf{CoarsestPartition}}
\def\partitioncoarsestm{\mathsf{CoarsestPartitionUpward}}
\def\topologicalorder{\mathsf{TopologicalSort}}

\def\adjustrngrolem{\mathsf{AdjustRoleLabels}}
\NewDocumentCommand{\rot}{O{45} O{1em} m}{\makebox[#2][l]{\rotatebox{#1}{#3}}}%

\begin{document}
\mainmatter              
\title{Dynamic Package Interfaces}
\subtitle{Extended Version}
\author{
Shahram Esmaeilsabzali\inst{1}\thanks{Shahram Esmaeilsabzali was at MPI-SWS when this work was done.} \and Rupak Majumdar\inst{2} \and Thomas Wies\inst{3} \and 
Damien Zufferey\inst{4}\thanks{Damien Zufferey was at IST Austria when this work was done.}
}

\institute{$^1$University of Waterloo\quad\quad $^2$MPI-SWS\quad\quad $^3$NYU \quad\quad $^4$MIT CSAIL\\
\email{sesmaeil@uwaterloo.ca},\email{rupak@mpi-sws.org}, \email{wies@cs.nyu.edu}, \email{zufferey@csail.mit.edu}
}

\maketitle              

\begin{abstract}
A hallmark of object-oriented programming is the ability to perform computation through a set of interacting objects.
A common manifestation of this style is the notion of a {\em package}, which groups a set of commonly used classes together.
A challenge in using a package is to ensure that a client follows the implicit protocol of the package when calling its methods.
Violations of the protocol can cause a runtime error or latent invariant violations. 
These protocols can extend across different, potentially unboundedly many, objects, and are specified
informally in the documentation. 
As a result, ensuring that a client does not violate the protocol is hard. 

We introduce {\em dynamic package interfaces (DPI)}, a formalism to explicitly capture the protocol of a 
package. 
The DPI of a package is a finite set of rules that together specify how any set of interacting 
objects of the package can evolve through method calls
and under what conditions an error can happen.
We have developed a dynamic tool that automatically computes an approximation of the DPI of a package, 
given a set of abstraction predicates. 
A key property of DPI is that the unbounded number of configurations of objects of 
a package are summarized finitely in an abstract domain.
This uses the observation that many packages behave monotonically: 
the semantics of a method call over a configuration does not essentially change 
if more objects are added to the configuration.
We have exploited monotonicity and have devised heuristics to obtain  succinct yet general DPIs.
We have used our tool to compute DPIs for several commonly used Java packages with
complex protocols, such as JDBC, HashSet, and ArrayList. 

\end{abstract}

\section{Introduction}

Modern object-oriented programming practice uses packages to encapsulate components, 
allowing programmers to use these packages through well-defined application programming interfaces (APIs).  
While programming languages such as Java and C\# provide a clear specification of the static APIs 
of a package in terms of classes and their (typed) methods, there is usually no specification of 
the implicit \emph{protocol} that constrains the temporal ordering of method calls on different objects. 
If the protocol is limited to a single object of a single class, it can be specified in form of a state 
machine whose states are the abstract states of the object and whose edges are the invocations 
of its methods \cite{Strom86:Type,Whaley02:Automatic,Alur05:Synthesis}.
For example, a lock object has two states: locked and unlocked. 
While in the unlocked (resp.\ locked) state, a call to the lock (resp.\ unlock) 
method takes it to the locked (resp.\ unlocked) state.
Any other method call results in an error.
The notion of state-machine interfaces has been studied extensively, and
there are many tools to generate interfaces using static or dynamic techniques \cite{Alur05:Synthesis,Henzinger05:Permissive,Pradel09:Automatic,DBLP:journals/ase/WasylkowskiZ11}.
However, existing notions of state machines on object states must be generalized when considering a package.
First, the internal state of an object should be considered in the context of the internal states of other objects; 
e.g., in the Java Database Connectivity (JDBC) package, a \texttt{Statement} object can execute safely only 
if its corresponding \texttt{Connection} object is open. 
Second, the execution of a method on an object can change the internal state of other objects in the environment; 
e.g., calling the \texttt{executeQuery} method on a JDBC \texttt{Statement} object closes its corresponding open \texttt{ResultSet} object. 
Finally, the protocol can constrain
the states and transitions of {\em unboundedly} many interacting objects; 
e.g., considering a collection object and its iterators, 
modifying the collection directly invalidates \emph{all} of its iterators.

The problem of generalizing interfaces from single to multiple objects has been studied recently \cite{Nguyen09:Graph,Pradel12:Static,Nanda05:Deriving}.
However, what is missing is a clear definition of what constitutes an interface in the presence of unboundedly many
objects on the heap.
Our first contribution is the introduction of \emph{dynamic package interface} (DPI), which allows to capture the protocol of a package in a succinct manner. 
The DPI of a package is a set of \emph{rules}, each of which specifies the effect of a method call on an object within an abstract \emph{configuration} of objects. 
An abstract configuration denotes an unbounded number of concrete configurations of objects from a package.
A rule has a \emph{source} and a \emph{destination} configuration, together 
with a \emph{mapping} that specifies how the objects in the source change to the objects in the destination.

Our first technical ingredient is a representation of abstract configurations using \emph{nested graphs} \cite{DBLP:conf/fossacs/WiesZH10}.
In a nested graph, a subgraph can be marked to be repeatable, and repetitions can be nested.
Nested graphs naturally represent unbounded heap configurations.
For example, Figure~\ref{fig:exampleng} shows a (two-level) nested graph
representing an open JDBC \texttt{Connection} object with its many corresponding closed \texttt{Statement} objects, 
each with many closed \texttt{ResultSet} objects.

Our second ingredient is an abstract semantics of Java-like languages over the domain of nested
graphs that is monotonic (in fact, the abstract transition system is \emph{well-structured} \cite{ACJT96}): 
if a method can be called in a ``smaller'' configuration, it can be also called in a ``larger'' configuration, with 
the resulting configurations maintaining the relationship. 
Monotonicity enables us 
to define the DPI rules of a package only over its {\em maximal} abstract configurations, 
letting each rule subsume infinitely many similar ``smaller'' rules. 
We prove that the set of maximal configurations has a finite representation, 
and thus the DPI of a package has a finite number of rules \cite{OOSemanticsTech}.

Our second contribution is a dynamic analysis technique to compute an approximation of the DPI of a package directly from the source code.
Our tool explores the usage scenarios of a package by running a \emph{universal client} that in each of its finite number of steps, 
nondeterministically, either creates a new object or invokes a method of an existing object.
Each step of the universal client results in a rule.
The universal client can end up computing hundreds or thousands of distinct rules, which makes the resulting DPI practically not useful.
The challenge is to generalize these rules to obtain a compact DPI by exploiting similarity.
Often, a pair of rules for the same method are incomparable only because their sources and destinations are slightly different. 
For example, in one rule for the \srct{close} method of the \srct{Statement} class, the source configuration  has closed \srct{ResultSet} objects but not an open one, and vice versa, another rule might have an open \srct{ResultSet} object but not closed ones.
It makes sense, however, to combine these two rules because the effect of the two rules are essentially the same: the \srct{Statement} object and its open \srct{ResultSet} object are closed.

We have devised three heuristics that generalize a set of explored rules into a smaller, more general set.
Our \emph{extrapolation} heuristic compares the configurations of different rules and deduces whether the configuration 
of a certain rule can be expanded by repeating part of it based on the repetitions observed in the configurations of other rules.
Our \emph{merge} heuristic combines two rules that are based on similar method invocations into one rule. 
Our \emph{exception isolation} heuristic combines two similar exception rules into one.
While merging is similar to the union of the two rules, exception isolation is closer to an intersection that isolates the root cause of an exception.
Our heuristics are all grounded in the monotonicity property of our abstract semantics.


We have used our tool to compute the DPIs of Java packages such as JDBC (26 rules), HashSet (16 rules), and ArrayList (15 rules).
The rules of these DPIs can be traced to their documentation, as well as to the programming errors discussed in online discussion groups.
Our tool more often than not computes the expected number of rules for these packages, but not all these rules are the most general ones.
Our tool never computes a rule that is not consistent with the behaviour of a package.
This is an indication that our heuristics are effective.

The remainder of the paper is organized as follows. 
Section \ref{sect:overview} presents an overview of DPI and how it is computed in our tool.
Section \ref{sect:preliminary} presents the notion of DPI formally.
Section \ref{sect:hgtong} presents the algorithm that converts a heap configuration into a nested object graph.
Section \ref{sect:exploration} describes how our tool explores the behaviour of a package and create rules.
Section \ref{sect:heuristics}, \ref{sect:merging}, and \ref{sect:isolation} describe our extrapolation, merging, and exception isolations heuristics, respectively.
Section \ref{sect:tool} discusses our implementation.
Section \ref{sect:experiences} presents our experimental results.
Section \ref{sect:conclusion} concludes our paper.

\section{Overview: Dynamic Package Interface of JDBC}\label{sect:overview}
We now explain through an example how our tool works to compute the DPI of a set of classes that are part of Java Database Connectivity (JDBC) package (more precisely the \srct{java.sql} package).

\subsection{JDBC}\label{overview:jdbc}

We consider four commonly-used classes of JDBC and their methods.
The \srct{Driver\-Manager} class allows to create a new connection to a database by invoking its static \srct{getConnection} method. 
The string parameter of the method specifies the type of database, its address, and the needed credentials to access it.
A \srct{Connection} object can serve multiple \srct{Statement} objects, each of which can be used to read or change the content of the database.
The \srct{createStatement} method of the \srct{Connection} class creates a new \srct{Statement} object.
SQL commands and queries are executed through the \srct{execute} and \srct{executeQuery} methods of the \srct{Statement} class.
Both methods accept a string argument that is an SQL statement.
The \srct{executeQuery} method returns a new \srct{ResultSet} object, which is a collection of rows retrieved from the database; the \srct{next} method can be used to traverse these rows. 
A \srct{Connection}, \srct{Statement}, or \srct{ResultSet} object is \emph{open} initially, but can be closed via their corresponding \srct{close} methods.
Invoking the \srct{executeQuery} method on a \srct{Statement} object causes an open \srct{ResultSet} object that references it to be closed, while creating a new open \srct{ResultSet} object.
If an object, or one of the objects that it references directly or transitively, is closed, invoking a non-\srct{close} method on it would raise an exception.

\subsection{System Input}
Besides the names of classes and the signatures of their methods, our tool receives a set of 
abstraction predicates over the attributes of the classes. 
A predicate is either \emph{scalar}, defined over the simple, non-reference attributes of the classes, or \emph{reference}, 
determining which objects of a class are related to which objects of another class via a certain reference attribute.
For simplicity, we assume these predicates are input by the user, but standard techniques based on Boolean methods 
and reference-valued fields in classes can be used to identify these predicates \cite{DBLP:journals/ase/WasylkowskiZ11}.

For example, in JDBC, the \srct{Statement} class has an \srct{active} attribute that determines whether it is open or not. 
This attribute is a unary scalar predicate, but in general a scalar predicate may read multiple fields from referenced objects.
We also use the \srct{applicationConnection} field of the \srct{Statement} class to define a reference predicate that determines which 
\srct{Statement} object points to which \srct{Connection} object. 
We define similar scalar predicates for the \srct{Connection} and \srct{ResultSet} classes, which determine whether their objects are open or closed.
We also define a reference predicate that determines which \srct{ResultSet} objects reference which \srct{Statement} objects.

We require that the set of reference attributes do not create a cycle when evaluated over objects: 
i.e., when objects are considered as nodes and the true valuations of reference attributes as directed edges, the resulting graph is acyclic.
This is necessary as some of our algorithms rely on computing the topological ordering of heap-related graphs.
This requirement can be relaxed: it is possible to allow the more general class of the depth-bounded graphs~\cite{OOSemanticsTech}.

\subsection{DPI Rules} 
\begin{figure}%
\centering

\input{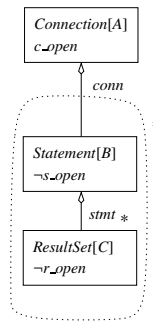}

\caption{A nested object graph.}
\label{fig:exampleng}
\end{figure}

The DPI of a package is a set of \emph{rules}, each of which represents a family of method calls.
A rule essentially specifies how a certain family of method calls change the shape of their corresponding heaps.
To obtain general yet concise rules, we have developed the domain of \emph{nested object graphs} to represent such heaps.
The nodes of a nested object graph represent objects and its edges represent references between the objects.
The nodes and edges of the graph are labelled according to the input scalar and reference abstraction predicates, respectively.
Furthermore, a subgraph of a nested object graph can be marked as repeatable, denoting that arbitrary-many sets of objects similar to the objects in the subgraph can exist in the heap.
Repetition can be nested, and hence the name ``nested object graph.'' 
As an example, the nested object graph in Figure \ref{fig:exampleng} represents a configuration of heap consisting of a \srct{Connection} object with unbounded many closed \srct{Statement} objects (possibly 0), each of which has unbounded many closed \srct{ResultSet} objects (possibly 0). 
Repetitions are specified via ``*'' next to nodes or subgraphs.
Node $C$, which represents the \srct{ResultSet} objects, is marked repeatable in a nested manner: Each group of repeatable \srct{ResultSet} objects is associated with a \srct{Statement} object, which itself is marked as repeatable via the ``*'' next to the subgraph specified by the dotted line.


Each rule has a source and a destination nested object graph, which correspond to the heaps before and after the method call.
A rule also has a source and a destination \emph{cast nested object graph}, each of which is a nested object graph some of whose nodes are labelled with roles, such as \emph{callee}, \emph{parameter\_0}, and \emph{new}, that specify the roles of objects in the method call.
The cast nested object graphs of a rule are meant to specify the objects in the heap that are directly involved in the method call, while the nested object graphs of the rule specify the entire heap affected by the method call.
A rule has an \emph{object mapping} (\emph{role mapping}) relation that specifies how, as a result of the method call, the objects represented by the nodes of the source nested object graph (correspondingly, source cast nested object graph) are transferred to the nodes of the destination nested object graph (correspondingly, destination cast nested object graph).
The mapping in each of these relations are annotated with multiplicity information that specify how many of the objects in the source node of a tuple are transferred to the destination node of the tuple: \emph{one} or \emph{many}.
Lastly, the object mapping and role mapping relations of a rule are derived from disjoint sets of Java objects: i.e., considering the underling method call related to a rule and the involved Java objects of the method call, each of the object is mapped either by the object mapping or role mapping of the rule, but not both.

\begin{figure}%
\centering

\subfigure[Role mappings.]{\input{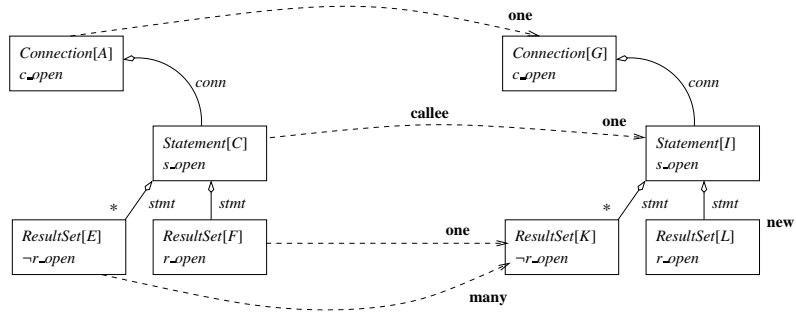}
\label{fig:rolemap2}}

\subfigure[Object mappings. The arrows over a nested subgraph denotes that all nodes of its source are mapped to their isomorphic nodes in the destination.]{\input{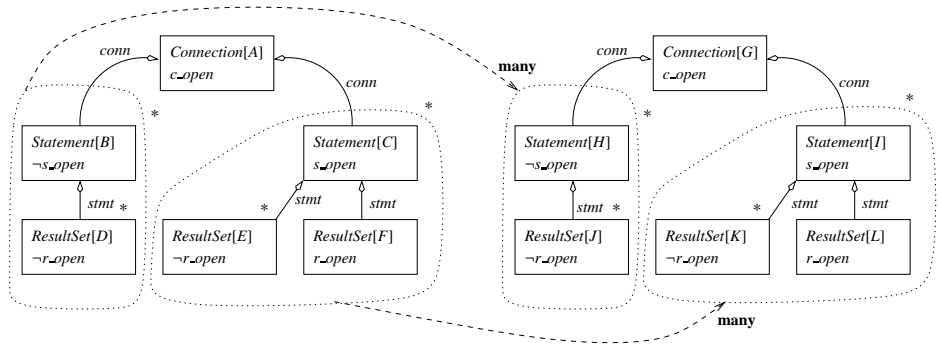}
\label{fig:objectmap2}}

\caption{The most general rule for \srct{executeQuery}, with no exception.}
\label{fig:rule2}
\end{figure}

As an example, Figure \ref{fig:rule2} shows the rule that our system computes for \srct{executeQuery} method calls that raise no exceptions.
The rule specifies that an open \srct{ResultSet} is closed when its corresponding \srct{Statement} object performs \srct{executeQuery}; instead, a new \srct{ResultSet} object is created.
Figure \ref{fig:rolemap2} specifies the role mapping of the rule, via dotted arrows that connect the nodes in the source cast nested object graph to the nodes in the destination cast nested object graph.
The ``callee'' and ``new'' labels determine the callee and the newly created objects, respectively.
Figure \ref{fig:objectmap2} specifies the object mapping of the rule via dotted arrows that, for the sake of brevity, connect the subgraphs of the nested object graphs.
While in this rule the object mapping does not specify any change in its corresponding objects, in general that is not the case.
Both nested object graphs and cast nested object graphs of the rule exhibit repetitions. 
In the case of the nested object graphs in Figure \ref{fig:objectmap2}, these repetitions are nested.
The left subgraph of the source nested object graph, for example, represents an arbitrary number (unbounded, possibly 0) of closed \srct{Statement} objects, each of which can have unbounded many closed \srct{ResultSet} objects.
It is this ability to express unbounded number of concrete heap configurations that allows us to compute general, yet concise interface rules.

While for a rule of a method call when it raises no exceptions, the more nodes and repetition that its nested object graphs have and the larger its mapping relations are the more general the rule would be (because it can capture more concrete method calls), for a rule for a method call with an exception that is not the case.
In fact, for such a rule it is desirable to have the smallest rule that isolates the real reason why the exception is raised.
As such, for an exception rule, we are only interested in its cast nested object graphs and their corresponding role mapping relation.
Furthermore, for exception rules, we use a ternary logic that assigns an unknown value ``*'' to a predicate of an object when the evaluation of the predicate does not affect whether the exception will be raised or not.
These characterizations of the most general rules for a method call are inspired by the monotonic semantics that we have developed for object-oriented programs.
For a safe method call, it should be possible to replicate its result in a context with more objects.
For a method call with an exception, there would not exist any context with more object that can avoid the exception.

\begin{figure}
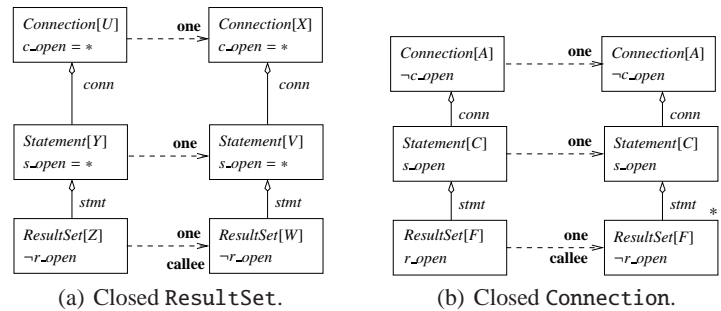
%
\centering

\subfigure[Closed \srct{ResultSet}.]{\input{figs/rolemaps3.pstex_t}
\label{fig:rolemaps3}}
\hspace{2em}
\subfigure[Closed \srct{Connection}.]{\input{figs/rolemap4.pstex_t}
\label{fig:rolemaps4}}

\caption{The two most general rules for \srct{next} with \srct{ResultSet not open} exception.}
\label{fig:rule3}
\end{figure}

Figure \ref{fig:rule3} shows the two rules that our tool computes for \srct{next} method when it raises the \srct{ResultSet not open} exception.
In Figure \ref{fig:rolemaps3}, the ``*'' values for the \srci{s\_open} and \srci{c\_open} predicates denote that regardless of whether the corresponding statement or connection objects of a \srct{Resultset} object are open or not, the method call over the \srct{Resultset} object raises the exception when it is closed.
Figure \ref{fig:rolemaps4} shows the case when the \srct{Resultset} object is actually open, but its corresponding \srct{Connection} object is not.
These two rules seem to point out succinctly the root cause of the bug discussed at an Apache forum.\footnote{\url{https://issues.apache.org/jira/browse/DERBY-5545}}

\subsection{From a Method Call to a Rule}
To compute the DPI of a package, our system explores the behaviour of the package through repeatedly invoking its methods and creating new rules.
A key step in computing a rule from a method call is to derive the necessary (cast) nested object graphs from different heaps.
In this section, we describe this through an example.

\begin{figure}%
\centering
\subfigure[Heap graph before method call.]{\input{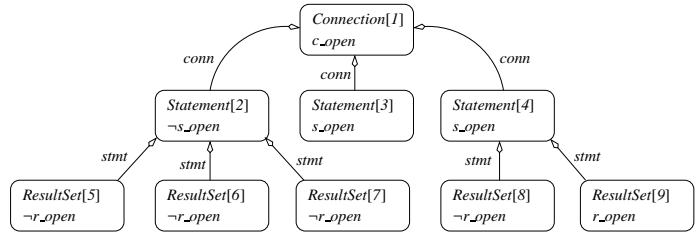}
\label{fig:og1}
}
\subfigure[Heap graph after method call.]{\input{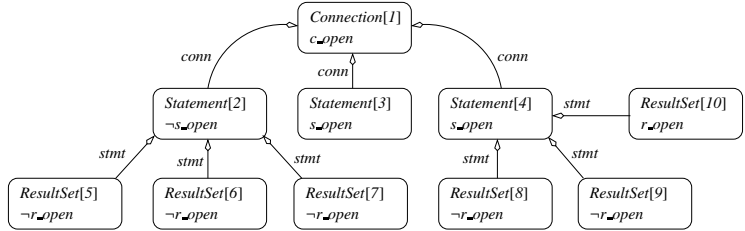}
\label{fig:og2}
}
\caption{Two heap graphs for invocation of \srct{executeQuery} on object 4.}
\label{fig:og}
\end{figure}

Our first step in computing the nested object graph of a heap is to turn the  heap into a directed labelled graph by using the input scalar and reference predicates. 
We call such a graph a \emph{heap graph}. 
Figure \ref{fig:og1} shows a heap graph corresponding to 9 JDBC objects.
The graph is created using three scalar predicates that determine whether a \srct{Connection}, \srct{Statement}, or \srct{ResultSet} object is open or not, together with two reference predicates that determine which \srct{Statement} objects reference which \srct{Connection} objects, and which \srct{ResultSet} objects reference which \srct{Statement} objects.
Each node of the graph is labelled with the name of its class, the evaluations of its scalar predicates, as well as a unique id that is enclosed inside a pair of brackets. 
Each edge of the graph is labelled with the name of its corresponding reference predicate.
Figure \ref{fig:og2} is another heap graph resulting from the invocation of method \srct{executeQuery} on the Java object that the node with id 4 in Figure \ref{fig:og1} represents.
The nodes with the same identifiers in the two objects graphs represent the same Java objects.

The second step is to reduce a heap graph to a nested object graph.
The idea is that if an object or a pattern for a set of interconnected objects appears more than once, then it can be marked as repeatable.
The reduction from a heap graph to an nested object graph can be considered as a bisimulation reduction: Two nodes in a heap graph are equivalent iff they have the same evaluations for their scalar predicates, and furthermore, they mimic one another by reaching equivalent nodes following their similar reference edges.
Figure \ref{fig:nog} shows two object graphs that our tool computes for the heap graphs in Figure \ref{fig:og}.
Repetition of a single node is denoted just by a ``*'' next to it.
Repetition of a subgraph (not shown in this figure) is denoted by a dotted line around the subgraph together with a ``*''; e.g., as in Figure \ref{fig:objectmap2}.
The nodes of the object graphs are graphically similar to heap graphs except that they are shown by solid rectangles and they are labelled with alphabetic ids.
As examples of repetition, node \srci{e} in Figure \ref{fig:nog1} is the equivalent class for the nodes  5, 6, and 7 in Figure \ref{fig:og1}, and node \srci{m} in Figure \ref{fig:nog2} is the equivalent class for the nodes 8 and 9 in Figure \ref{fig:nog2}.


\begin{figure}%
\centering
\subfigure[Nested object graph corresponding to heap graph in Figure \ref{fig:og1}.]{\input{figs/nog1.pstex_t}
\label{fig:nog1}
}
\hspace{2em}
\subfigure[Nested object graph corresponding to heap graph in Figure \ref{fig:og2}.]{\input{figs/nog2.pstex_t}
\label{fig:nog2}
}
\caption{Two nested object graphs.}
\label{fig:nog}
\end{figure}

To compute a rule, first, the set of objects that are relevant in computing the rule are determined.
These are used to create the nested object graphs of the rule.
A subset of these objects that are directly involved in the method call are used to create the cast nested object graphs of the rule, as well as its role mapping.
The object mapping of the rule deals with the rest of objects that are not mapped by its role mapping.

\subsection{Computation Stages}\label{overview:computation}

Creating a rule from a method call provides an abstract representation of the method call, but this abstraction is not nearly enough to create a succinct interface of the package: We could end up creating hundreds or thousands of  rules.
Algorithm \ref{alg:main} outlines the main steps that our tool performs to compute the DPI of a package.
Next, we describe these steps briefly.
More details about each step appears in its corresponding section that is mentioned inside comments in Algorithm \ref{alg:main}.

\begin{algorithm}[]
\SetAlgoLined
\SetKwInOut{Input}{input}\SetKwInOut{Output}{output}
\KwIn{A set of classes and methods and a set of abstraction predicates} 
\KwResult{A set of general rules, $\rules$, each of which represents a family of method calls} 

$\mathit{\rules = \emptyset}$\;

\tcc*[h]{Section \ref{sect:hgtong} and \ref{sect:exploration}}

\While{$\mathit{\neg Threshold}$} {\nllabel{line:startexplore}

{Pick a snapshot, a concrete Java object, execute one of its methods\;}

{Compute, $\mathit{r}$, the corresponding rule of the method call\; \nllabel{line:mainderiverule}}

\lIf{there is no $\mathit{r' \in \rules}$ that ``covers'' $r$}{$\mathit{\rules = \rules \cup \{r\}}$\;}

}
{Remove any $\mathit{r \in \rules}$ that is ``covered'' by another rule\;\nllabel{line:endexplore}}

\tcc*[h]{End Section \ref{sect:hgtong} and \ref{sect:exploration} .}

{Extrapolate $r\in \rules$ using $r' \in \rules$, when possible; prune rules that are covered by $r$\;\tcc*[h]{Section \ref{sect:heuristics}.}\nllabel{line:widengeneral}}  

{Merge all pairs of mergeable rules in $\rules$\;\tcc*[h]{Section \ref{sect:merging}.}\nllabel{line:mergegeneral}}

{Isolate all pairs of similar exception rules in $\rules$\;\tcc*[h]{Section \ref{sect:isolation}.}\nllabel{line:isolationexception}}
\caption{$\mathit{ComputeDPI}$.} \label{alg:main}
\end{algorithm}

\paragraph{Exploration Stage} 
Lines \ref{line:startexplore} - \ref{line:endexplore} specify the main steps in exploring the behaviour of a package.
Secion~\ref{sect:exploration} explains how a rule is computed from the execution of a method.
Using a repository, our tool keeps track of the rules that it explores.
For each computed rule, $r$, it checks whether there already exists a rule $r'$ that \emph{covers} $r$, roughly meaning that the object graphs, role graphs, object mapping, and role mapping of $r$ can be all in a way simulated by the corresponding elements of $r'$.
If such an $r'$ exists, $r$ is \emph{redundant} and is not stored. 
The system continues its exploration until a maximum number of redundant method invocations is encountered; e.g., in our experiments with JDBC we set this threshold to 1200.
After this initial phase of exploration, to achieve a good coverage of the behaviour of the package, our system also ensures that all possible method calls on all objects of all rules in the repository are executed and their corresponding rules, if non-redundant, are stored in the repository.
Lastly, all redundant rules are pruned from the repository.

\paragraph{Extrapolation Stage}
In order to obtain a DPI with a small number of rules, our tool generalizes rules so that one generalized rule covers many other rules.
In the absence of such general rules, many incomparable rules can be explored and stored, making a DPI too large to be of any practical use.
Sometimes a rule could have covered many other rules if certain nodes in its source and/or destination (cast) nested graphs were marked as repeatable. 
Our tool uses an \emph{extrapolation heuristic} to mark such nodes as repeatable using the information in the graphs of other rules.

To identify opportunities for extrapolation, our tool looks for \emph{deficient} nodes in a (cast) nested object graph: A node is deficient if it is not repeatable and belongs to a pair in one of the the two mapping relations of a rule, and the other node in the pair is repeatable.
Our hypothesis is that a deficient node is not repeatable because our exploration has not managed to produce enough objects of a that type.
As an example, if we consider the graphs in Figure \ref{fig:nog} as the corresponding nested object graphs of a rule, then \srci{f} and \srci{g}, which would be both mapped to node \srci{m}, are deficient nodes.
For a deficient node, our system explores all other rules in its repository checking for a source or a destination object graph into which the corresponding object graph of the deficient node can be \emph{embedded} according to a subgraph isomorphism relation.
If according to the embedding relation the corresponding node of the deficient node in the other graph is repeatable, then the deficient node will be marked as repeatable too. 
In our example, our tool can find an embedding relation that would allow to extrapolate node \srci{f}, but it cannot extrapolate node \srci{g}, because in JDBC each \srct{Statement} object cannot have more than one open \srct{ResultSet} object.

After the exploration stage, we apply our extrapolation heuristic to all rules.
Once all possible extrapolation have been performed, our tool checks for redundant rules and removes them.
While the extrapolation stage could prune a substantial number of rules, there could still exist a large number of rules in a DPI; e.g., hundreds of rules for JDBC.
The reason is that different rules for the same method might have explored different instances of heaps that have incomparable sets of objects, and there are various exception cases.
To further reduce the number of the rules of a DPI, we have developed two heuristics -- the \emph{merge} heuristic and the \emph{exception isolation} heuristic. Each of the heuristics combines a set of related rules into one.

\paragraph{Merging}

For a pair of rules whose role mappings are similar and over isomorphic cast nested object graphs, the merge heuristic essentially first computes their union and then performs a reduction over the resulting source and destination object graphs of the resulting rule.
This reduction can be considered as a bisimulation reduction except that two nodes could be equivalent even if one has some incoming edges that the other one does not have. 
This is as opposed to the kind of reduction that we described for reducing heap graphs to object graphs earlier.
This reduction is in the spirit of \emph{downward closed} graphs, where a nested object graph not only represents all heap instances arising from the repetition of its repeatable subgraphs, but also represents all heap instances arising from its nested object \emph{sub}graphs -- hence the term ``downward closed''.
The reduction favours repetition over non-repetition when combining nodes. 
Finally, the role mapping and object mapping of the resulting rule are adjusted according to the reduction.
As an example, assuming that the nested object graphs in Figure \ref{fig:nog} belong to a rule, then node $c$ in Figure \ref{fig:nog1}, for instance, would be mapped to node $C$ in Figure \ref{fig:objectmap2} during the merge operation.

\paragraph{Exception Isolation}

While the first merging heuristic corresponds to the union of a set of rules, the second heuristic corresponds to the intersection of a set of rules.
For a pair of rules whose role mappings are isomorphic when their scalar abstraction predicates are not considered, this heuristic essentially combines the corresponding nodes of the cast nested object graphs of the two rules and merges equivalent nodes via a ternary logic.
If the value of a predicate in two merged nodes are different, the unknown value, denote by ``*'', is chosen. 
Figure \ref{fig:rule3} shows the two rules that our tool computes for the \srct{next} method on a \srct{Resultset} object when it raises the \srct{ResultSet not open} exception.
The ``*'' values for the \srci{s\_open} and \srci{c\_open} predicates denote that regardless of whether the corresponding \srct{Statement} or \srct{Connection} objects of a result set object are open or not, the method call raises the exception when the result set object is closed.

%

\section{DPI Formally}\label{sect:preliminary}
\paragraph{Graph Definitions.}
A directed multigraph is a tuple $G = (V,E,s,t,l)$, where $V$ is a set of nodes, $E$ is a set of edges, $s: E\rightarrow V$ is the \emph{edge source} function, $t: E\rightarrow V$ is the \emph{edge destination} function, and $l: E \rightarrow \edgelabels$ is the \emph{edge labelling} that assigns a string label to each edge. 
Node $u_2 \in G.V$ is \emph{reachable from} a node $u_1 \in G.V$, or $u_1$ \emph{reaches} $u_2$, if a sequence of edges connect $u_1$ to $u_2$.
By $\reachablefrom{G}{u}$, we denote the set of all nodes that are reachable from $u$ plus $u$ itself; similarly, by $\reachingto{G}{u}$, we denote the set of all nodes that reach $u$, plus $u$ itself.
By $\reachablespanning{G}{u}$, we denote the set of all nodes that are reachable from $u$, assuming that for each $e \in G.E$, we add $e'$ to $G.E$ such that $s(e') = t(e)$, $t(e) = s(e')$, and $l(e') = l(e)$ (i.e., assuming that $G$ is a undirected graph).
We also extend these notation to work with a set of nodes; e.g., $\reachablefrom{G}{U} = \bigcup_{u\in U} \reachablefrom{G}{u}$.
By $\reachablefromto{G}{u}$, we denote the set of all nodes that that reach $u$, plus those that are reached from $u$ and those that reach $u$: i.e., $\reachablefromto{G}{u} = \reachablefrom{G}{\reachingto{G}{u}}$.
For a set of nodes $U \subseteq G.V$, by $\subgraph{G}{U}$, the \emph{induced subgraph} of $G$ over $U$ is a directed multigraph that is the same as $G$ but its elements are restricted to $U$.

A graph $H$ is \emph{subgraph isomorphic} to $G$ if there exists two injective  mappings  $k_v: H.V\rightarrow G.V$ and $k_e: H.E\rightarrow G.E$ such that:
$$
H' =(\bigcup_{v \in H.V}k_v(v), \bigcup_{e \in H.E}k_e(e), \bigcup_{(e,v) \in H.s}(k_e(e),k(v)), \bigcup_{(e,v) \in H.t}(k_e(e),k(v)), \bigcup_{(e,b) \in H.l}(k_e(e),b)~)
$$ 
\noindent is an induced subgraph of $G$ over $H'.V$; we call $k_v$ and $k_e$, respectively, the \emph{node} and \emph{edge} \emph{isomorphism mapping} of $H$ to $G$.
A graph $H$ is \emph{graph isomorphic} to $G$ if $|H.V| = |G.V|$, $|H.E| = |G.E|$, and $H$ is subgraph isomorphic to $G$. 

Two distinct nodes, $u_1$ and $u_2$, of $G$ are \emph{coinciding} if for each $e_1 \in G.E$ such that $s(e_1) =u_1$ there exists $e_2 \in G.E$ such that $s(e_2) =u_2$, $t(e_1) = t(e_2)$, and $l(e_1) = l(e_2)$, and furthermore, vice versa: for each edge whose source is $u_2$ there is a corresponding edges whose source is $u_1$ and the two edges have the same target and label.
Two distinct nodes, $u_1$ and $u_2$, of $G$ are \emph{downward consistent} if they are coinciding, and $\subgraph{G}{\reachingto{G}{u_1}}$ and $\subgraph{G}{\reachingto{G}{u_2}}$ are isomorphic.

\paragraph{Modelling Heap.}
A Java class is represented as a tuple, $C = (\cname,\atts,\creators,\modifiers)$, where $\cname$ is the \emph{name} of the class, $\atts$ is its set of \emph{attributes}, $\creators$ is its set of \emph{creator} methods, each of which is either a constructor or a static method that returns a new object, and $\modifiers$ is its set of \emph{modifiers} methods, each of which can be invoked on an object of the class, changing its attributes.
An attribute is either \emph{primitive}, meaning that its type is a simple type, or \emph{reference}, meaning that its type is a class. 
A method can have a set of formal parameters and a return value, each of which can be a class.
A \emph{package}, $\packagej$, is a set of classes.

A Java object, also called a \emph{concrete object}, is represented äs a tuple, $o = (\coid,\coclass)$, where $\coid$ is its unique \emph{object id} and $\coclass$ is its corresponding class.
A concrete object, $o$, can \emph{reach} another concrete object, $o'$, if by following a sequence of reference attributes starting from $o$, $o'$ is reached.
A \emph{snapshot}, $\snapshot$, is a set of concrete objects. 
A \emph{role} is a tuple $l = (\co,\rolename)$, where $\co$ is a concrete object and $\rolename$ is the \emph{role name}, which is a string representing the responsibility of the object in a method call, e.g., ``callee'', ``return'', ``new'', or ``param1''.
A method call is represented as an \emph{invocation}, which is a tuple, $\invoc = (\method,\exceptionname,\srcsp,\destsp,\roles)$, where  $\method$ is the method, $\exceptionname$ is the name of an exception if the method call raises the exception and empty otherwise, $\srcsp$ is the \emph{source snapshot}, which is the snapshot before the method call, $\destsp$ is the \emph{destination snapshot}, which is the snapshot after the method call, and $\roles$ is a set of roles corresponding to the method call.

A \emph{scalar predicate} over an object is an abstraction predicate over its primitive attributes and, possibly, the primitive attributes of the objects that are reachable from it. 
A \emph{reference predicate} is an abstraction predicate over a \emph{source} object, a \emph{destination} object, and a reference attribute of the source object.
Its value is true if the source object references the destination object through its reference attribute, and is false otherwise.
These predicates are defined over the classes of a package and are evaluated with respect to the objects of a snapshot.
We assume that these predicates are defined such that they can always be evaluated: i.e., it is never the case that a scalar predicate cannot be evaluated because a certain object that is assumed to be reachable is not reachable.

An \emph{abstract object} is a tuple $\ao = (\co,\preds)$ where $\co$ is a concrete object and $\preds$ is the evaluation of its corresponding scalar predicates.
Two abstract objects, $\ao$ and $\ao'$, are \emph{equivalent}, denoted by $\ao \equivalentao \ao'$, if their corresponding predicates have the same valuations.

A \emph{heap graph} is a directed, acyclic multigraph, $\hg = (\hgnodes,\hgedges,\hgsrc,\hgdest,\hgedgelabel)$, whose nodes are abstract objects and whose edges are labelled by reference attributes. 
Given a snapshot $\sp$ and a set of scalar and reference predicates, the \emph{underlying heap graph} of $\sp$, denoted by $\hg(\sp)$, is a heap graph whose nodes are the corresponding abstract objects of the concrete objects in $\sp$ and whose edges correspond to the true valuations of the reference predicates over the objects in $\sp$; the labels of the edges correspond to the names of their corresponding reference attributes.
By construction of an underlying heap graph, no two edges with the same source node have the same label.
We assume that the reference predicates are defined such that for any snapshot $\sp$, $\hg(\sp)$ is acyclic.

A \emph{nested abstract object} is a tuple $\nao = (\naoid,\ao,\plural,\injective)$, where $\naoid$ is its unique \emph{id}, $\ao$ is its \emph{representative} abstract object, 
$\plural$ is its \emph{plural} flag, and $\injective$ is its \emph{injective} flag.
If either $\plural$ or $\injective$ is true, then $\nao$ represents more than one equivalent abstract objects, otherwise $\nao$ is \emph{singular} and represents a single abstract object. 
These two flags are used to denote the two kinds of equivalent abstract objects that a nested  abstract object can represent.
Intuitively, if $\nao.\plural$ is true, then $\nao$ represents a group of equivalent abstract objects, represented by $\nao.\ao$, that point to the same abstract objects via their same reference attributes. 
Intuitively, if $\nao.\injective$ is true, then $\nao$ represents a group of equivalent abstract objects, represented by $\nao.\ao$, that pairwise disagree at least on the destination of one of their reference attributes.
By analogy to entity relationship modelling, the plural flag represents a many-to-one relationship and the injective flag represents \emph{many} one-to-one relationships.
A nested abstract object ${\nao}_1$ is \emph{equivalent} to a nested abstract object ${\nao}_2$, denoted by ${\nao}_1 \equivalentnao {\nao}_2$, if: (i) ${\nao}_1.\ao \equivalentao {\nao}_2.\ao$, (ii) ${\nao}_1.\plural \equiv {\nao}_1.\plural$, and (iii) ${\nao}_1.\injective \equiv {\nao}_1.\injective$.
A nested abstract object ${\nao}_1$ is \emph{smaller} than nested abstract object ${\nao}_2$, denoted by ${\nao}_1 \smallernao {\nao}_2$, if: (i) ${\nao}_1.\ao \equivalentao {\nao}_2.\ao$, and (ii) neither ${\nao}_1.\plural$ nor ${\nao}_1.\injective$ is true, but either ${\nao}_1.\plural$ or ${\nao}_1.\injective$ is true.
A nested abstract object ${\nao}_1$ is \emph{covered} by a nested abstract object ${\nao}_2$, denoted by ${\nao}_1 \smallereqnao {\nao}_2$, if either ${\nao}_1 \equivalentnao {\nao}_2$ or ${\nao}_1 \smallernao {\nao}_2$.
Given a pair of nested abstract objects, ${\nao}_1$ and ${\nao}_2$, the \emph{renesting} of ${\nao}_1$ with ${\nao}_2$, denoted by $\renest{{\nao}_1}{{\nao}_2}$, modifies ${\nao}_1$ such that ${\nao}_1.\plural = {\nao}_2.\plural$ and ${\nao}_1.\injective = {\nao}_2.\injective$.

A \emph{nested object graph} is a directed, acyclic graph, $\ng = (\ngnodes,\ngedges,\ngsrc,\ngdest,\ngedgelabel)$, whose nodes are nested abstract objects and whose edges are labelled by reference attributes. 
We use nested object graphs as a means to generalize heap graphs.
By the definition of this generalization, which will be presented in Section \ref{sect:exploration}, there is not more than one edge with the same label between two nodes of a nested object graph;
also, for each edge $({\nao}_1,{\nao}_2)$ of a nested object graph if both ${\nao}_1.\plural$ and ${\nao}_1.\injective$ are false, then it should be the case that both ${\nao}_2.\plural$ and ${\nao}_2.\injective$ are also false.
A \emph{nested object graph} ${\ng}_1$ is \emph{covered} by a nested object graph ${\ng}_2$, denoted by ${\ng}_1 \smallereqng {\ng}_2$, if: (i) ${\ng}_1$ is subgraph isomorphic to ${\ng}_2$ when the two graphs are considered as simple graphs whose nodes have unique labels and whose edges are the same as the original, and (ii) for any pair of isomorphic nodes, ${\nao}_1 \in {\ng}_1.\ngnodes$ and ${\nao}_2 \in {\ng}_2.\ngnodes$, ${\nao}_1 \smallereqnao {\nao}_2$.
The \emph{corresponding} nested object graph of a heap graph $\hg$, denoted by $\hgtong{\hg}$, is the same as $\hg$ except that each node, $\ao \in \hg.\hgnodes$, is replaced with a nested abstract object $\nao = (\mathit{newid},\ao,\mathit{false},\mathit{false})$, where $\mathit{newid}$ is a unique id.

A \emph{cast nested object graph} is a tuple $\rng = (\rngnodes,\rngedges,\rngsrc,\rngdest,\rngedgelabel,\rngrole)$, where $(\rngnodes,\rngedges,\rngsrc,\rngdest,\rngedgelabel)$ is a nested object graph and $\rngrole$ is a \emph{role labelling} function that relates a \emph{role}, derived from an invocation, to a nested abstract object.
This function is not surjective as it only labels the objects that are directly labelled by the roles of the corresponding invocation of the graph.

\paragraph{Rules.}

A \emph{mapping} is a tuple $m = (\srcnao,\destnao,\mult)$ that relates a \emph{source} nested abstract object $\srcnao$ to a \emph{destination} nested abstract object $\destnao$ via a \emph{multiplicity} value $\mult$, which is either \emph{one} or \emph{many}.
A ``one'' multiplicity means that exactly one abstract object represented by $\srcnao.\ao$ is mapped to one abstract object represented by $\destnao.\ao$.
A ``many'' multiplicity means more than one such abstract objects are mapped.
A singular source nested abstract object can be mapped  only via a one multiplicity; similarly, a destination nested abstract object can be mapped to via a one multiplicity.
A mapping $m$ is \emph{covered} by a mapping $m'$, denoted by $m \smallereqmap m'$ if: (i) $m.\srcnao \smallereqnao m'.\srcnao$, (ii) $m.\destnao \smallereqnao m'.\destnao$, and (iii) it is not the case that $m.\mult = \textrm{many}$ and $m'.\mult = \textrm{one}$.

A \emph{rule} is a tuple $r = (\method,\exceptionname,\srcng,\destng,\srcrng,\destrng,\mapng,\maprng)$, where:
\begin{itemize}
\item $\method$ is the corresponding method of the rule;

\item $\exceptionname$ is either the name of an exception of the method or is empty; 

\item $\srcng$ and $\destng$ are the \emph{source} and \emph{destination} nested object graphs of the rule, respectively;

\item $\srcrng$ and $\destrng$ are the \emph{source} and \emph{destination} cast nested object graphs of the rule, respectively;

\item $\mapng \subseteq \srcng.\ngnodes \times \destng.\ngnodes \times \set{one,many}$ is the \emph{object mapping} relation, which is a set of mappings such that any node in $\srcng.\ngnodes$ or in $\destng.\ngnodes$ is part of at least one mapping; and

\item $\maprng \subseteq \srcrng.\ngnodes \rightarrow (\destrng.\ngnodes \times \set{one,many})$ is the \emph{role mapping} relation, which is a set of  mappings such that any node in $\srcrng.\ngnodes$ or in $\destrng.\ngnodes$ is part of at least one mapping, except for a node whose role is ``new'' or ``return''.

\end{itemize}

A \emph{dynamic package interface}~(DPI) is a set of rules.

\section{From Heap Graphs to Nested Object Graphs} \label{sect:hgtong}

In Algorithm \ref{alg:main}, a key step is to compute a rule from a method invocation (line \ref{line:mainderiverule}).
The challenging aspect of this step is how to compute the necessary graphs of a rule from the corresponding heap graphs corresponding to the heap before and the heap after the method invocation.
Given a heap graph, $\hgtongalg$ in Algorithm \ref{alg:hgtong} computes a nested object graph that is structurally a minimization of the heap graph, similar to the bisimulation reduction of a transition system or DFA minimization.  
The difference, however, is that the resulting nested object graph embodies also information about the repetition patterns in the heap graph.
Next, we describe this algorithm in more detail.

\begin{algorithm}[]
\SetAlgoLined
\SetKwInOut{Input}{input}\SetKwInOut{Output}{output}
\KwIn{A heap graph, $\hg$} 
\KwResult{A nested object graph, $\ng$}
{$\ng = \hgtong{\hg}$\;\nllabel{line:hgtong}}
{$\objectsorder = \topologicalorder(\ng)$\;\nllabel{line:toposort}}
\ForEach{$\naos \in \objectsorder$ visited in the topological sort order}
{\nllabel{line:reduce1loop}
{$\partition = \findsimilars(\ng,\naos)$\;\nllabel{line:parition1}}
{$\ng = {\lumpng}(\ng,\partition)$\;\nllabel{line:lump1}}
}

{$\coarsestpartition = {\partitioncoarsest}(\ng,\ng.\ngnodes)$\;\nllabel{line:partition2}}

{\Return{${\lumpngfinal}(\ng,\coarsestpartition)$}\;\nllabel{line:returnlumped}}

\caption{$\hgtongalg$ Algorithm.} \label{alg:hgtong}
\end{algorithm} 

In the first step of the algorithm, the input heap graph is converted to its corresponding nested object graph (line \ref{line:hgtong}).
In the second step, the nodes of the resulting nested object graph are sorted according to a topological sort order that puts different sets of incomparable nodes in their corresponding equivalence sets; the result is a sequence of sets of nodes, with the nodes with no incoming edges in the first set and the nodes with no outgoing edges in the last set; it is stored in $\objectsorder$ (line \ref{line:toposort}).
The next step is to traverse through these sets and reduce each by combining their similar nodes (the loop on line \ref{line:reduce1loop}).
Function $\findsimilars$, on line \ref{line:parition1}, finds the set of nested abstract objects that can be summarized into one; function $\lumpng$, on line \ref{line:lump1}, lumps the graph by replacing such nodes with one representative node.
Next, we describe these two functions.

Function $\findsimilars$ takes a set of nested abstract objects, $\naos$, and partitions them to a set of sets of nested abstract objects each of which is a maximal set of pairwise downward consistent nodes, such that for each pair of nodes, $u_1$ and $u_2$, in the set, $\subgraph{\ng}{\reachingto{\ng}{u_1}}$ and $\subgraph{\ng}{\reachingto{\ng}{u_2}}$ are isomorphic via equivalent nested abstract objects.
Each set $\block1\in\partition$, e.g., the set of closed \srct{ResultSet} objects pointing to the same \srct{Statement} object, represents a set of concrete objects that are pointing to the same concrete objects (because nodes are processed in the topological sort order and because the nodes in ${\block}_1$ are pairwise coinciding) and are pointed by similar objects (because their corresponding downward subgraphs are isomorphic).
Function $\lumpng$ takes a ${\block}_1 \in \partition$ and removes all nodes of ${\block}_1$ together with their corresponding edges from $\ng$, except one node, which is randomly chosen and we refer to as ${\nao}_{\mathit{rep}}$. 
If $|{\block}_1|>1$, ${\ng}_{\mathit{rep}}.\plural$ is set to true, indicating that ${\ng}_{\mathit{rep}}$ represents ``many'' objects.

After the loop on line \ref{line:reduce1loop} terminates, a summarized nested object graph is obtained.
However, this graph can be further summarized.
As an example, let us consider a set of closed \srct{ResultSet} concrete objects that point to the same concrete \srct{Statement} object.
The operations in the loop on line \ref{line:reduce1loop} lumps such \srct{ResultSet} concrete objects into one, but if there are two such sets of \srct{ResultSet} concrete objects that point to two different concrete \srct{Statement} objects, the result would be two lumped \srct{ResultSet} nested abstract objects that point to the same lumped \srct{Statement} nested abstract object. 
The two \srct{ResultSet} nested abstract objects, however, should also be lumped, as they are essentially the same and already point to the same \srct{Statement} object.

Function $\partitioncoarsest$, on line \ref{line:partition2}, identifies opportunities for such lumpings.
It is essentially a partition refinement algorithm, akin to DFA minimization or bisimulation reduction algorithms, that starts with an initial partition of the set of all nodes of $\ng$ and refines this partition until the partition cannot be further refined. 
The initial partition consists of the set of sets of equivalent nested abstract objects.
A block, ${\block}_2$, of a partition, ${\partition}_2$, can be refined if: (i) some of the nodes in ${\block}_2$ have incoming edges with a certain label, while the others do not have such incoming edges; or (ii) some of the nodes in $\block2$ have outgoing edges with a certain label to another block, while the others do not have such outgoing edges.
In either case, such a block is partitioned into two blocks.
The rational for this refinement is to distinguish between nested abstract objects that are pointed to or point to different types of nested abstract objects.
For example, the refinement distinguishes between \srct{Statement} objects that are pointed by open \srct{ResultSet} objects and those that are not.
Finally, function $\lumpngfinal$ lumps the nested object graph, $\ng$, according to the partitions of $\coarsestpartition$, and returns the result (line \ref{line:returnlumped}).
Function $\lumpngfinal$ is the same as $\lumpng$, except that when choosing a representative nested abstract object, ${\nao}_{\mathit{rep}}$, of a block, ${\block}_2 \in \coarsestpartition$, if $|{\block}_2|>1$ and ${\nao}_{\mathit{rep}}.\plural = \mathit{false}$, then ${\nao}_{\mathit{rep}}.\injective$ will be marked as true.
Setting ${\nao}_{\mathit{rep}}.\injective$ to true models the many one-to-one relationship; e.g., it is used to model the case when many open individual \srct{ResultSet} objects point to their corresponding \srct{Statement} objects.

Algorithm $\hgtongalg$ returns also a mapping that specifies how the node of $\hg$ are mapped to the nodes of $\ng$.

\paragraph{Nesting Level}
Lastly, we describe an alternative method to represent the nesting structure of a nested object graph; we have used this method in our formal abstract semantics for OO programs \cite{OOSemanticsTech}. 
This method naturally describes the \emph{nesting level} of each node of a graph via a number, which can be either zero or a positive number.
The nesting level of zero for a node denotes no repetition.
A positive nesting level for a node specifies repetition, but the scope of repetition also depends on the nesting level of the neighbouring nodes of the node.
Adjacent nodes with the same nesting level that is greater than zero together denote the repetition of the subgraph that they represent.
Repetition can be nested through nodes that have edges to nodes with less nesting levels.
Based on the observation that the nesting level of a source node that has an edge to destination node cannot be less than the nesting level of the destination node, we have developed a simple algorithm to assign nesting levels to the nodes of nested object graph.
First, we sort the nodes of the nodes of the graph according to the opposite topological sort order in a list: i.e., the list starts with the nodes with no outgoing edges and ends with the nodes with no incoming edges.
We then process the nodes in the list as follows.
For a nested abstract object, $\nao$, let $\mathit{max}$ be the maximum nesting level of the immediate nodes that it can reach with its outgoing edges.
If $\nao.\plural$ is true, then its nesting level would be $\mathit{max}+1$, if $\nao.\injective$ is true, then it is $\mathit{max}$, if both $\nao.\plural$ and $\nao.\injective$ are false then it 0.
It can be shown that this assignment of nesting levels captures the intended repetition structure of a nested object graph.
As an example, this scheme assigns 0, 1, and 2 to the nodes $A$, $B$, $C$ of the nested object graph in Figure \ref{fig:exampleng}, respectively.

\section{Exploration Stage}\label{sect:exploration}

Algorithm \ref{alg:explore} shows how our tool explores the behaviour of a package, using a notion of \emph{universal client}. 
The universal client consists of a while loop (line \ref{line:universalloop}) that continues to execute the methods of the classes of a package until a maximum number of redundant rules are visited.
The exploration is random in that the method that is to be executed, the snapshot on which the method will be executed, the callee object, and the actual parameters are all chosen randomly. 
When a new explored method invocation is not \emph{covered} by any of the already-explored rules in $\rules$ (line \ref{line:newrule}), it is added to $\rules$; the new snapshot created as a result of the method call is also added to the set of already-explored snapshots, $\snapshots$.
The formal definition of rule coverage is presented at the end of this section, but intuitively, a rule $r$ is covered by a rule $r'$, if each of the elements of $r$ has a corresponding element in the elements of $r'$.

After an initial exploration of a package, our tool \emph{completes} it exploration by executing all modifier methods, on all distinct objects of all snapshots.
But before doing that it prunes all redundant rules in $\rules$(line \ref{line:prune1}).
Function $\completifyalg$, on line \ref{line:complete}, takes the set of already-generated rules, $\rules$, and for each $r \in \rules$ executes all possible modifier method calls over the corresponding concrete objects of the nested abstract objects of $r.\srcng$ and $r.\destng$.
(Note that a nested abstract object is related to an abstract object, which in turn is related to a concrete object and a snapshot.)
For each such invocation, if a new non-redundant rule is resulted, it is added to the $\rules$.
As observed by others~\cite{Dallmeier10:Generating}, trying to complete what has already been explored could improve the coverage of exploration.
At the end, the set of rules, $\rules$, is returned after being pruned of redundant rules.

Next, we describe the key steps of this algorithm in more detail.

\begin{algorithm}[]
\SetAlgoLined
\SetKwInOut{Input}{input}\SetKwInOut{Output}{output}
\KwIn{A set of classes, their methods together with a set of scalar and reference predicates over the classes} 
\KwResult{An distinct set of initial rules} 

$\rules= \emptyset$\;
$\snapshots = \{{\snapshot}_0\}$, where ${\snapshot}_0$ is empty\;

\While{$\mathit{redundants < maxRedundant}$} {\nllabel{line:universalloop}

Pick $\method$, a method of a class $C$ randomly\;
Pick $\snapshot \in \snapshots$ randomly\;
Pick $\actualparams$, the actual object parameters for $\method$ randomly from $\snapshot$\;
\If{$\method \in C.\modifiers$}
{Pick a callee object, $\calleeobj$, from $\snapshot$ randomly\;}
Execute $\method$ over $\calleeobj$ using $\actualparams$, and then derive invocation tuple $\invoc$\;
{$r = {\createrulealg}(\invoc)$\;\nllabel{line:createrule}}
$\mathit{redundantFlag = false}$\;

\ForEach{$\mathit{r'\in \rules}$}
{
\If{$r \smallereqrule r'$}
{\nllabel{line:newrule}
$\mathit{redundants++}$\; 
$\mathit{redundantFlag = true}$\; 
\textbf{break}\;
}
}

\If{$\mathit{\neg redundantFlag}$}
{
$\mathit{\rules = \rules\cup \set{r}}$\;
$\mathit{\snapshots = \snapshots \cup \set{\invoc.\destsp}}$\;
}

}

{$\mathit{\rules} = {\prunealg}(\mathit{\rules})$\;\nllabel{line:prune1}}
{$\mathit{\rules} = {\completifyalg}(\mathit{\rules})$\;\nllabel{line:complete}}

\Return{$\prunealg(\rules)$}\;

\caption{$\explorealg$ Algorithm.} \label{alg:explore}
\end{algorithm}

\subsection{Creating A Rule}
The call to function $\createrulealg$ on line \ref{line:createrule} transforms an input method invocation, $\invoc$, to a generalized rule, $r$.
The first two elements of $r$, $r.\method$ and $r.\exceptionname$, are simply $\invoc.\method$ and $\invoc.\exceptionname$, respectively.
We next describe how other elements of $r$ are computed.

To compute the source and destination nested object graphs of $r$, the $\hgtongalg$ algorithm in Algorithm \ref{alg:hgtong} is employed.
In our analysis, the source and destination nested object graphs of a rule are meant to include \emph{all} objects that are involved in the corresponding method call of the rule, as well as those objects that could be possibly affected by the method call.
Our hypothesis is that these objects can be characterized as the ones that could be reachable from or could reach to the objects in $\invoc.\roles$, plus the ones that reach such objects.
At below, by $\raos$, we denote the domain of $\invoc.\roles$.
The nested object graphs are then:

\begin{tabular}{ l l l }

$r.\srcng$ & $=$ &${\hgtongalg}({\hg}(\reachablespanning{\invoc.\srcsp,\raos})$; and\\
$r.\destng$ & $=$ &${\hgtongalg}({\hg}(\reachablespanning{\invoc.\destsp,\raos})$.

\end{tabular}

\noindent (We can use $\raos$ objects both for computing the source and destination nested abstract object graphs because the unique ids of objects identify them in different snapshots.)

In our analysis, the source and destination cast nested abstract object graphs of a rule are meant to include the objects that are directly involved in the method call.
These objects are the one that are reachable from or reach to the objects in $\raos$.
As such, the nested object graph components of $r.\srcrng$ and $r.\destrng$ are computed using  a variation of $\hgtongalg$ that ensures that each element of $\raos$ belongs to a singleton partition, otherwise, the dependencies between various objects that are cast will be lost through lumping of similar nodes.
This can be achieved by modifying the $\findsimilars$ and $\partitioncoarsest$ functions each to accept a parameter that specifies the nodes that each requires its own block.
The role labelling components of $r.\srcrng$ and $r.\destrng$ can be derived from $\invoc.\roles$ by using the mapping information generated by Algorithm $\hgtongalg$ that relates concrete objects to nested abstract objects.

Lastly, $r.\mapng$ is computed by keeping track of how abstract objects that were used to create $r.\srcng$ and $r.\destng$, but were not used in the creation of $r.\srcrng$ and $r.\destrng$, are mapped from a source nested abstract object to a destination nested abstract object.
The computation of $r.\maprng$ is similar to the computation of $r.\mapng$, considering only the abstract objects that were used to create $r.\srcrng$ and $r.\destrng$.

Lastly, we present the formal definition of covering relation between two rules.
\paragraph{Rule Comparison}

Rule $r$ is \emph{covered} by rule $r'$, denoted by $r \smallereqrule r'$, if:
\begin{itemize}
\item $r.\method = r'.\method$,
\item $r.\exceptionname = r'.\exceptionname$,
\item $r.\srcng \smallereqng r'.\srcng$, according to a node subgraph isomorphism mapping $N_1 \subseteq r.\srcng.\ngnodes \rightarrow r'.\srcng.\ngnodes$,
\item $r.\srcng \smallereqng r'.\srcng$, according to a node subgraph isomorphism mapping $N_2 \subseteq r.\destng.\ngnodes \rightarrow r'.\destng.\ngnodes$,
\item $(r.\srcrng.\rngnodes,r.\srcrng.\rngedges,r.\srcrng.\rngsrc,r.\srcrng.\rngdest,r.\srcrng.\rngedgelabel) \smallereqng$ \\ $(r'.\srcrng.\rngnodes,r'.\srcrng.\rngedges,r'.\srcrng.\rngsrc,r'.\srcrng.\rngdest,r'.\srcrng.\rngedgelabel)$, according to a node subgraph isomorphism mapping $R_1 \subseteq r.\srcrng.\rngnodes \rightarrow r'.\srcrng.\rngnodes$,  
\item $(r.\destrng.\rngnodes,r.\destrng.\rngedges,r.\destrng.\rngsrc,r.\destrng.\rngdest,r.\destrng.\rngedgelabel) \smallereqng$ \\ $(r'.\destrng.\rngnodes,r'.\destrng.\rngedges,r'.\destrng.\rngsrc,r'.\destrng.\rngdest,r'.\destrng.\rngedgelabel)$, according to a node subgraph isomorphism mapping $R_2 \subseteq r.\destrng.\rngnodes \rightarrow r'.\destrng.\rngnodes$,
\item for any $m \in r.\mapng$, there exists a mapping $m'\in r'.\mapng$ such that:
\begin{itemize} 
\item $(m.\srcnao,m'.\srcnao) \in N_1$,
\item $(m.\destnao,m'.\destnao) \in N_2$, and 
\item $m \smallereqmap m'$;

\end{itemize}
 
\item for any $m \in r.\maprng$, there exists a mapping $m' \in r'.\maprng$ such that:
\begin{itemize}

\item $(m.\srcnao,m'.\srcnao) \in R_1$,
\item $(m.\destnao,m'.\destnao) \in R_2$, 
\item $m \smallereqmap m'$,

\item $r.\srcrng.\rngrole^{-1}(m.\srcnao) = r'.\srcrng.\rngrole^{-1}(m'.\srcnao)$, and
\item $r.\destng.\rngrole^{-1}(m.\destnao) = r'.\destng.\rngrole^{-1}(m'.\destnao)$.

\end{itemize}

\end{itemize}

\section{A Heuristic for Rule Generalization}\label{sect:heuristics}

The result of running the $\explorealg$ in Algorithm \ref{alg:explore} is a set of rules, each of which is a generalization of a particular method call.
While such a generalization may be able to identify repetition pattern over some parts of the heap, it may not identify repetition for other parts.
One way to further generalize a rule is to try to \emph{extrapolate} the nodes in its nested object graphs and cast nested object graphs: for the singular nodes in these graphs, try to change their $\plural$ and $\injective$ properties to true, whenever it is possible.

We have developed a heuristic for extrapolating the nodes of the graphs in a rule.
The extrapolation opportunities are identified by checking the anomalies in the object mapping and role mapping of a rule: if one element of a mapping is singleton but not the other, then our heuristic tries to find another rule that has a similar, but more general version of this mapping, in which case the graphs in the original rule can be generalized based on the nodes in the graphs of the other rule.
Next, we describe our heuristic for the case where we deal with the nested object graphs of a rule and its object mapping.
The heuristic for cast nested object graphs of a rule is similar.

\subsection{Extrapolation of Deficit Nested Abstract Objects}

Given a rule, $r$, a nested abstract object, $\nao$, is \emph{deficit}, if $\nao$ is singular, and there exists a mapping $m \in r.\mapng$ such that: either $m.\srcnao = \nao$ and $m.\destnao$ is not singular, or $m.\destnao = \nao$ and $m.\srcnao$ is not singular.
For example, if $m.\srcnao$ is a singular, open \srct{ResultSet} object, while 
$m.\destnao$ is a non-singular closed \srct{ResultSet} object, then $m.\srcnao$ is deficit.

Our heuristic is based on the hypothesis that a nested abstract object, $\nao$, is deficit as a result of the universal client having not explored certain use cases of a package.
If there is a nested object graph of another rule, either its source or destination nested object graph, that has a node, ${\nao}_{p}$, such that $\nao.\ao \equiv {\nao}_{p}.\ao$ and ${\nao}_{p}$ is not singular, then our hypothesis could be somewhat validated: there is no inherent reason for ${\nao}$ to be singular; it is perhaps singular because of insufficient exploration.
However, this observation does not take into account that ${\nao}$ and ${\nao}_{p}$ could belong to two nested object graphs with different structures.
A nested abstract object may have to be only singular in one structure, but need not be singular in another structure. 
Thus, the extrapolation of the deficit nested abstract object, ${\nao}$, is allowed only if 

$$
\subgraph{G}{\reachablefromto{G}{\nao}} \smallereqng \subgraph{H}{\reachablefromto{H}{{\nao}_{p}}},
$$
\noindent where $G$ and $H$ are the nested object graphs that $\nao$ and ${\nao}_{p}$ belong to, respectively.

The extrapolation of a node in a graph, however, should be consistent in that the resulting graph should not have an edge whose destination is a smaller nested object graph than its source; e.g., a singular open \srct{ResultSet} object should not point to a non-singular open \srct{Statement} object.
As such, in our heuristic, we extrapolate a group of objects together.
Considering the node isomorphism mapping, $k_v$, between the nodes of the above subgraphs, for each $({\nao}_G,{\nao}_H) \in k_v$, we extrapolate ${\nao}_G$ via $\renest{{\nao}_G}{{\nao}_H}$.
In our experience, we have observed that this collective extrapolation always precludes creating any ill-formed nested object graph, although the collective extrapolation only applies to a subset of the nodes of $G$ and $H$.

This heuristic is also applied to the nodes of cast nested abstract object graphs of a rule by identifying the deficit nodes in the role mappings of the rule.
The only difference is that a node that is labelled by the role labelling function of its corresponding cast nested object graph cannot be extrapolated.
Those objects are inherently singular.

In our tool, we apply the extrapolation heuristic to all rules in an arbitrary order.
After this step, we also check one more time to see whether further extrapolation can be performed using the elements of the rules that have already been extrapolated.

\subsection{Adjusting the Multiplicity of a Mapping}
One source of inconsistency that could arise as a result of applying our heuristic would be the creation of a mapping whose multiplicity is ``one'' while its source or destination nested abstract object has been extrapolated from a singular nested abstract object to a non-singular one.
We can fix a certain class of such inconsistencies by changing these multiplicities to ``many''; namely, if both the source and the destination of a mapping $m$ are non-singular and $m.\mult = \mathit{one}$, and $m.\srcnao$ is not the source of any other mapping, then $m.\mult$ is set to $\mathit{many}$.
For other cases, it is not certain that by changing $m.\mult$ to ``many'', we will not change the semantics of the rule.
In our experiences, the above fix captures almost all the needed adjustments.

\section{Rule Merging}\label{sect:merging}

The extrapolation heuristic can result in more general rules, which in turn render many explored rules as redundant.
However, there could still exist may distinct rules for a DPI.
The merging of a pair of rules results a new rule that can be considered as a 
summary of the union of the two rules.
Thus, the two rules are replaced with one.
The rationale to allow for such a merging is rooted in the abstract semantics that we have developed for OO programs over \emph{well-structured transition systems} \cite{OOSemanticsTech}: Given a method invocation over a nested object graph, it can be replicated in a \emph{larger} nested object graph, where the notion of ``larger'' is similar to the covering relation between nested object graphs in this paper.
Based on this property, in our merging algorithm, we rely on the assumption that by taking a rule, $r$, and adding objects of a similar rule, $r'$, to $r$, we only generalize $r$ further, but do not introduce a behaviour that is not observable in the package.
Of course, this assumption does not always hold in a dynamic analysis, but in our experiences, it always held.


Two rules, $r$ and $r'$, are \emph{mergeable}, denoted by $\mergeable{r}{r'}$, if the following conditions hold:
\begin{itemize}

\item $r.\method = r'.\method$,
\item $r.\exceptionname = r'.\exceptionname$, and
\item $\roleconsistent{r}{r'}$,

\end{itemize}

\noindent where two rules, $r$ and $r'$, are \emph{role consistent}, denoted by $\roleconsistent{r}{r'}$, if:
\begin{itemize}

\item $\subgraph{r.\srcrng}{\reachablefrom{r.\srcrng}{\range{r.\srcrng.\rngrole}}}$ and \\
$\subgraph{r.\srcrng}{\reachablefrom{r.\srcrng}{\range{r.\srcrng.\rngrole}}}$  are isomorphic according to a node graph isomorphism mapping $M_1 \subseteq r.\srcrng.\rngnodes \rightarrow r'.\srcrng.\rngnodes$, in which if $({\nao}_1,{\nao}_2)\in M_1$, then ${\nao}_1.\ao \equivalentao {\nao}_2.\ao$;

\item $\subgraph{r.\destrng}{\reachablefrom{r.\destrng}{\range{r.\destrng.\rngrole}}}$ and \\
$\subgraph{r.\destrng}{\reachablefrom{r.\destrng}{\range{r.\destrng.\rngrole}}}$  are isomorphic according to a node graph isomorphism mapping $M_2 \subseteq r.\destrng.\rngnodes \rightarrow r'.\destrng.\rngnodes$, in which if $({\nao}_1,{\nao}_2)\in M_2$, then ${\nao}_1.\ao \equivalentao {\nao}_2.\ao$; and
 
\item For any role mapping $m \in r.\maprng$ such that $m.\srcnao \in \range{r.\srcrng.\rngrole}$, there exists a role mapping $m' \in r'.\maprng$ such that:
\begin{itemize}
\item $(m.\srcnao,m'.\srcnao)\in M_1$;
\item $(m.\destnao,m'.\destnao)\in M_2$; and
\item $r.\maprng^{-1}(m.\srcnao) = r'.\maprng^{-1}(m'.\srcnao)$; i.e., the two nodes have the same role label.
\end{itemize}
\item And similarly, for any $m'.\srcnao \in \range{r'.\srcrng.\rngrole}$, there exists a matching role mapping $m \in r.\maprng$.

\end{itemize}
\noindent Essentially, two rules are mergeable if they are based on similar invocations. 
The role consistency criteria checks the similarity of objects that have role labels and ignores the rest of nested abstract objects in the cast nested object graphs of the two rules that are likely not to be common in all related invocations.
Based on our observation that method calls can be generalized over \emph{downward-closed} heaps \cite{OOSemanticsTech}, we characterize these non-consequential nested abstract objects as those that can make a rule ``larger'' but not essentially different.

Algorithm \ref{alg:mergeall} shows the algorithm that merges all mergeable rules of a given set of rules. 
When a pair of rules are merged, the first rule is replaced with the result of merging (line \ref{line:mergecall}), while the second is removed from the result (line \ref{line:mergedelete}).
The key function is $\mergealg$, which merges two rules into one.

\begin{algorithm}[]
\SetAlgoLined
\SetKwInOut{Input}{input}\SetKwInOut{Output}{output}
\KwIn{A set of rules, $\rules$} 
\KwResult{A set of merged rules, $\rules$} 

$\mathit{copyofRules} = \rules$\;

\ForEach{$r \in \mathit{copyofRules}$}
{
\lIf{$r \notin \rules$}
{\textbf{continue}\;}

\ForEach{$r'\in \rules$ such that $r' \neq r$}
{
\If{$\mergeable{r}{r'}$}
{

{$r = {\mergealg}(r,r')$\;\nllabel{line:mergecall}}
{$\rules = \rules - \set{r'}$\;\nllabel{line:mergedelete}}
}
}
}
\Return{$\mathit{\rules}$}\;

\caption{$\mergeallalg$ Algorithm.} \label{alg:mergeall}
\end{algorithm}

The $\mergealg$ algorithm itself is essentially based on two algorithms that combine a pair of cast nested object graphs ($\combinerngs$ algorithm) and nested object graphs ($\combinengs$ algorithm). 
Before describing $\mergealg$, we first describe $\combinerngs$; $\combinengs$ is similar to $\combinerngs$.

Algorithm \ref{alg:combinerngs} presents the $\combinerngs$ algorithm. 
It accepts a pair of cast nested object graphs, $\rng$ and ${\rng}'$, together with an isomorphism mapping $M$ resulting from checking that the two graphs are mergeable, and creates a new cast nested object graph, $\mrng$.
The steps between lines \ref{line:startsimple} and \ref{line:endsimple} compute the union of nodes and edges of $\rng$ and ${\rng}'$, while removing those nodes of ${\rng}'$ that has an isomorphic node in ${\rng}$.
In the process of removing these nodes, the corresponding node of ${\mrng}$ is renested.
The result is a nested object graph whose elements are stored in $\rng$.
Once these nodes and edges are computed, a coarsest partition of these nodes is computed.
This computation is the same as the computation on line \ref{line:partition2} in Algorithm \ref{alg:hgtong}, except that two nodes can belong to the same block of partition even if their incoming edges do not match.
This partitioning results in a reduction of the graph that embodies the semantics of downward-closed heaps \cite{OOSemanticsTech}.
Based on this partition, function $\lumpngm$, on line \ref{line:lumpm}, reduces the corresponding nested object graph of $\mrng$.
This algorithm is similar to $\lumpngfinal$ used in line \ref{line:returnlumped} of Algorithm \ref{alg:hgtong}, except that when choosing a representative nested abstract object for a block, a non-singular node is chosen if there exists; if only singular objects exist, then the representative will be singular.
Lastly, based on the partition, the role labelling of $\mrng$ needs to be adjusted (line \ref{line:adjustrolem}).

Applying $\combinerngs$ to a pair of graphs could create an inconsistent graph that has edges whose sources are singular but their destinations are not; this happens because of renesting that is done in the process.
For such edges, which were rarely observed in our experiences, there is a function that adjust their source nodes to have the same $\plural$ and $\injective$ properties as their destinations.

\begin{algorithm}[]
\SetAlgoLined
\SetKwInOut{Input}{input}\SetKwInOut{Output}{output}
\KwIn{A pair of cast nested object graphs, ${\rng}$ and ${\rng}'$, and an isomorphism mapping, $M \subseteq {\rng}.\rngnodes \times {\rng}'.\rngnodes$} 
\KwResult{A new nested object graph, $\mrng$, which is a summary of the input graphs}

{$\mrng.\rngnodes = r.\rngnodes \cup r'.\rngnodes - \range{M}$\;\nllabel{line:startsimple}}

\ForEach{$({\nao}_1,{\nao}_2) \in M$}
{
$\mrng.\rngnodes = \mrng.\rngnodes \cup \set{\renest{{\nao}_1}{{\nao}_2}}$\;
}
 
$\mrng.\rngedges = {\rng}.\rngedges$;~~$\mrng.\rngsrc = {\rng}.\rngsrc$~~$\mrng.\rngdest = {\rng}.\rngdest$\;
\ForEach{$e \in {\rng}'.\rngedges$}
{
\If{${\rng}'.{\rngsrc}(e) \in \range{M} \wedge {\rng}'.{\rngdest}(e) \in \range{M}$}
{\textbf{continue}\;}
$\mrng.\rngedges = \mrng.\rngedges \cup \set{e}$\;

\lIf{${\rng}'.{\rngsrc}(e) \in \range{M}$}{$\mrng.\rngsrc = \mrng.\rngsrc \cup \set{(e,M^{-1}({\rng}'.{\rngsrc}(e)))}$\;}

\lIf{${\rng}'.{\rngdest}(e) \in \range{M}$}{$\mrng.\rngdest = \mrng.\rngdest \cup \set{(e,M^{-1}({\rng}'.{\rngdest}(e)))}$\;}

}

{$\mrng.\rngedgelabel = {\rng}.\rngedgelabel \cup {\rng}'.\rngedgelabel$\;\nllabel{line:endsimple}}

{$\coarsestpartitionm = {\partitioncoarsestm}(\mrng,\mrng.\ngnodes)$\;\nllabel{line:partitionm}}

{$\mrng = {\lumpngm}(\mrng,\coarsestpartitionm)$\;\nllabel{line:lumpm}}

{$\mrng.\rngrole = {\adjustrngrolem}(r.\rngrole,\partitioncoarsestm)$\;\nllabel{line:adjustrolem}}

\Return{$\mrng$}\;

\caption{$\combinerngs$ Algorithm.} \label{alg:combinerngs}
\end{algorithm}

Function $\combinengs$ is the same as $\combinerngs$ except that it does not accept $M$ as an input, and that the union of the corresponding nested object graphs of $\rng$ and ${\rng}'$ is simply the union of their elements.

Algorithm \ref{alg:merge} presents function $\mergealg$, which uses $\combinengs$ and $\combinerngs$ to merge a pair of rules, $r$ and $r'$. 
Lines \ref{line:srcrngmerge} and \ref{line:destrngmerge} combine the source cast nested object graphs of $r$ and $r'$.
Function $\combinemaps$, on line \ref{line:adjustrngmap}, first computes the union of the role mappings of the two rules, and then adjusts them to the nodes of the cast abstract object graphs of $\mrule$; $L_1$ is a mapping that specifies how each node of the cast nested object graphs of the original rules are mapped to the nodes of the cast nested object graphs of $\mrule$.
The next three lines are similar, but deal with nested object graphs. 

Function $\combinemaps$ may come across two maps between the same pair of nested object graphs, with one having \emph{one} and one with \emph{many} multiplicity. 
In such a case, only the mapping with the \emph{many} multiplicity is kept.
Lastly, if, a nested abstract object is the destination of two maps, it cannot be singular; similarly, a singular nested abstract object cannot be the sole destination of a non-singular nested abstract object, neither the multiplicity of such a mapping can be \emph{one}. These anomalies are all adjusted.

\begin{algorithm}[]
\SetAlgoLined
\SetKwInOut{Input}{input}\SetKwInOut{Output}{output}
\KwIn{A pair of mergeable rules, $r$ and $r'$} 
\KwResult{A new rule, $\mrule$, that is the merge of input rules} 
$\mrule.\method = r.\method$ and $\mrule.\exceptionname = r.\exceptionname$\;

{$\mrule.\srcrng$ = $\combinerngs(r.\srcrng,r'.\srcrng,M_1)$\;\nllabel{line:srcrngmerge}}
{$\mrule.\destrng$ = $\combinerngs(r.\destrng,r'.\destrng,M_2)$\;\nllabel{line:destrngmerge}}
{$\mrule.\maprng$ = $\combinemaps(r.\mapng,r'.\mapng,L_1)$\;\nllabel{line:adjustrngmap}}
{$\mrule.\srcng$ = $\combinengs(r.\srcng,r'.\srcng)$\;\nllabel{line:srcngmerge2}}
{$\mrule.\destng$ = $\combinengs(r.\destng,r'.\destng)$\;\nllabel{line:destngmerge2}}
{$\mrule.\mapng$ = $\combinemaps(r.\maprng,r'.\maprng,L_2)$\;\nllabel{line:adjustrngmap2}}
\Return{$\mrule$}\;

\caption{$\mergealg$ Algorithm.} \label{alg:merge}
\end{algorithm}

\section{Exception Isolation}\label{sect:isolation}

While the $\mergeallalg$ algorithm is effective in deriving the most general rules of a package for the method calls that do not raise exceptions, it is not well-suited for method calls with exceptions.
The reason is twofold.
First, often when an exception is raised the states of objects of a rule do not change.
Thus adding more nodes to the cast nested object graphs and nested object graphs of a rule, through merging, does not make the rule any more general.
Second, for rules with exceptions, it is often not important to determine the fates of objects \emph{after} the method call, but rather it is important to determine what was especial about the states of objects \emph{before} the method call.

We have developed a method for summarizing rules with exceptions, called \emph{exception isolation}, which, similar to the merge process in the previous section, combines a set of related rules into one.
Furthermore, it addresses the above two considerations.
First, we only consider the cast nested object graphs of the rules when isolating a certain exception.
Second, we use a three-value logic, including true, false, and an unknown value ``*'', that can combine the objects of the same class that have different scalar predicate values.
It works as follows.
For a pair of isomorphic objects identified in a a pair of cast nested object graphs, if a certain predicate has different values for the two objects, the value of the predicate is set to ``*''.
The unknown values for predicates isolate the root cause of when a method can raise a certain exception.
When checking for isomorphism, however, we need to make sure that an object, $\nao$, will not be paired with an object whose predicates is not the same as $\nao$, while there indeed exists another object whose predicates conform with the ones of $\nao$; e.g., an open \srct{ResultSet} object should not be matched with a closed \srct{ResultSet} if it can be matched with an open one.
Thus, to avoid premature combination of objects with different predicates, we first apply the merge algorithm on rules with exceptions, in order to have larger graphs that decrease the chance of combining objects prematurely.

Our three-valued combination scheme, however, could sometimes create too coarse an isolation for a set of rules. 
For example, if we further isolate the rules in Figure \ref{fig:rolemaps3} and Figure \ref{fig:rolemaps4} by combining them into one, then the resulting rule would prescribe that the \srct{next} method always raises an exception.
To avoid such over-isolations, our isolation algorithm can be tuned to pair two callee objects of two rules as isomorphic only if they are equivalent.
Our tool has an option that specifies whether callee objects can be combined in three-value logic or not.
One way to decide whether to apply three-value combination to callee objects or not, is to first allow this type of combination, and check whether an overlap happens. 
If it does, then try the isolation process without three-value combination of callee objects, otherwise accept the resulting set of rules.
While for JDBC package, we chose not to combine the pairs of isomorphic callee objects in the three-value logic, for the Array-Iterator package, we chose to combine them in three-value logic.

\section{System}\label{sect:tool}

\begin{figure}
\centering
\input{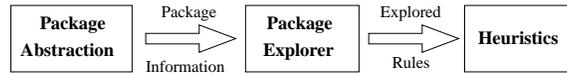}
\caption{The main components of the system.}
\label{fig:architecture}
\end{figure}

Figure \ref{fig:architecture} shows the high-level architecture of our system, which is implemented in Java.
The arrow between the components of the system specify the high-level input communicated between these components.

The \emph{Package Abstraction} component provides the abstract information about the package that our tool uses to compute the DPI of the package.
It is a programmatic way to provide an input to our system.
It consists of a set of classes whose methods specify the classes of the package under study, the methods of these classes, and the valuations of the abstraction predicates of the objects of the package.
These classes basically use Java reflection to present the aforementioned information about the package under study.
Furthermore, there are classes that provide the actual parameters for the method calls of the universal client; these actual parameters have random values.
While we manually create these classes, but many of them can be automatically generated based on inputs from a user; e.g., the names of chosen classes,  abstraction predicates over their attributes, etc.

The \emph{Package Explorer} component essentially implements our exploration algorithm described in Section \ref{sect:exploration}.
To implement a notion of snapshot whose objects can be accessed throughout the exploration, we need to be able to obtain a copy of an object of a snapshot on demand.
To achieve this, for each snapshot of objects, our tool maintains the corresponding trace of method calls that resulted in the snapshot.
To call a method of an object of a snapshot, our tool recreates the entire snapshot by replaying its corresponding trace. 
(Cloning or saving an object, in general, would not work, as not all classes implement these methods.)
A recreated snapshot has similar objects as the original snapshot, assuming that, as far as the abstraction predicates are concerned, method calls are deterministic.
To relate the objects in a snapshot to the objects in its replayed copy, we use a notion of \emph{logical id} for each of the objects of the snapshots; 
objects that have the same logical ids are treated as copy of one another.

In our implementation of the exploration algorithm, as opposed to the algorithms in Section \ref{sect:exploration}, we use the nested abstract object graphs of a rule to represent its cast nested abstract object graphs as well.
These cast nested abstract object graphs are in a sense an \emph{unfolding} of their corresponding nested object graphs, as described in \cite{OOSemanticsTech}.

To ensure that our exploration does not prematurely identify a certain kind of object as singular in a rule, we use a repetitive object creation scheme in our exploration: If a creator method is chosen to be executed, we invoke the method $n>1$ number of times consecutively, and only after that compute the rule with respect to the snapshot before consecutive method calls and the snapshot after that. 
To avoid undesired redundant method calls, a method is not called more than once on the same object of a snapshot; similarly, our system nondeterministically chooses not to execute a method over an object of a snapshot if the last method call in the snapshot is the execution of the same method, possibly on a different object.

Lastly, the \emph{Heuristic} component implements the algorithms in Section \ref{sect:heuristics}, \ref{sect:merging}, and \ref{sect:isolation}. We use the graph data structures in JGraphT library to implement our graph algorithms.

\subsection{Limitations} While we expect our tool to work in a straightforward manner on packages that solely work on heap (e.g., Java collections), for packages that work with external devices, the Package Abstraction part of this system would be more complex. 
Furthermore, for such packages, the effect of our ``replay'' mechanism should be taken into account.
For example, if a trace of method cause a certain port to be bound, the naive replay of the same trace would not obtain a new copy of the object of interest.
Instead, a different port during the replay should be used.
These limitations are not unique to our approach, but are inherent to dynamic approaches.

\section{Experiences}\label{sect:experiences}
We have computed the DPI of three Java packages using our tool: \srct{JDBC}, \srct{ArrayList}, and \srct{HashSet}. 
While our tool usually identifies the right number of rules for the DPI of a package, some of these rules could be in principle more generalized. 
The converse, however, has never happened in our experiments: i.e., a rule for a packaged computed by our tool always corresponded to the actual behaviour of the package.

Table \ref{tab:results} shows the results of running our tool for each of these packages.
The measurements for each package are for the average of five runs on a dual-core CPU Windows 7 desktop machine with 8 GB of RAM. 
In all our experiments, we have set JVM options to use 5120MB of physical memory and to avoid raising a Garbage Collection exception, because of the lack of progress in computation.
For each package, Table \ref{tab:results} presents the time taken and the number of rules at each stage of the computation of a DPI, namely after the exploration phase, after the extrapolation phase, and after the merge phase.
We use the line numbers of Algorithm \ref{alg:main}, in Section \ref{sect:overview}, to indicate the stages that these measurements have been performed; e.g., line number \ref{line:endexplore} denotes a measurement after the computation at line \ref{line:endexplore} has concluded.

\begin{table}[t]
    \centering
    \caption{\label{tab:results} Duration and number of rules after different stages in computing DPIs of three packages.
    Information, except for the last column, correspond to average values of five runs.} 
    \begin{tabular}{cccccccccc}

&& \rot[43]{Exploration} & \rot[43]{Extrapolation} & \rot[43]{Merging} & \rot[43]{Isolation} &  \rot[43]{Exploration} & \rot[43]{Extrapolation} & \rot[43]{Merging} & \rot[43]{Isolation}  \\
\cline{1-10}
\multicolumn{1}{ |c| }{Package}& \multicolumn{1}{ |c| }{Threshold \#}  & \multicolumn{4}{ |c| }{Time (min:sec)} &  \multicolumn{4}{ |c| }{\#Rules}\\ \cline{1-10}

\multicolumn{1}{ |l| }{\srct{ArrayList}} & \multicolumn{1}{ |c| }{200000} &  \multicolumn{1}{ |c| }{010:37}  & \multicolumn{1}{ |c| }{000:03} & \multicolumn{1}{ |c| }{000:00} & \multicolumn{1}{ |c| }{000:00} & \multicolumn{1}{ |c| }{572} & \multicolumn{1}{ |c| }{299} & \multicolumn{1}{ |c| }{29} & \multicolumn{1}{ |c| }{15 (once 14)} \\
\cline{1-10}

\multicolumn{1}{ |l| }{\srct{HashSet}} & \multicolumn{1}{ |c| }{200000} &  \multicolumn{1}{ |c| }{168:26}  & \multicolumn{1}{ |c| }{000:23} & \multicolumn{1}{ |c| }{000:01} & \multicolumn{1}{ |c| }{000:00} & \multicolumn{1}{ |c| }{1140} & \multicolumn{1}{ |c| }{503} & \multicolumn{1}{ |c| }{34} & \multicolumn{1}{ |c| }{16} \\
\cline{1-10}

\multicolumn{1}{ |l| }{\srct{JDBC}} & \multicolumn{1}{ |c| }{1200} &  \multicolumn{1}{ |c| }{032:01}  & \multicolumn{1}{ |c| }{000:57} & \multicolumn{1}{ |c| }{000:05} & \multicolumn{1}{ |c| }{000:00} & \multicolumn{1}{ |c| }{2465} & \multicolumn{1}{ |c| }{2370} & \multicolumn{1}{ |c| }{29} & \multicolumn{1}{ |c| }{26 (twice 25) } \\
\cline{1-10}

\end{tabular}
    
\end{table}


\paragraph{JDBC.} In Section \ref{sect:overview}, we have already presented some of the rules of the DPI of JDBC.
In our experiments, the universal client connects to a local Apache Derby database. 
We use a key-value table that is manipulated through INSERT, DELETE, and SELECT SQL commands with random values, via JDBC.
We are thus assuming that the DPI of the JDBC package is independent of the schema of databases that can be connected via JDBC, which is justified by our interest in determining the relationship of interacting objects of a package, and not its interaction with external components.
Increasing the threshold value to a value bigger than 1200 could cause out-of-memory exceptions in our system. 
Our tool computes 26 rules in three out of five runs; in the other two runs, it computes 25 rules.
The missing rule in both cases is the rule for \srct{close} method call over an open \srct{ResultSet} object that references a closed \srct{Statement} object that in turn references a closed \srct{Statement} object.

\paragraph{ArrayList.} 
We consider two classes of \srct{ArrayList}: \srct{Array} and its internal class \srct{Itr}, which implements Java \srct{Iterator}. 
Besides the creator methods for these classes, we consider the \srct{Add} method of \srct{Array}, and the \srct{next} and \srct{remove} methods of 
\srct{Itr}.
We provide a reference predicate, $iter\_of$, to the system, which determines which \srct{Itr} object belongs to which \srct{Array} object.
We provide four scalar predicates to the system: $\mathit{empty \equiv size>0}$, which determines whether an \srct{Array} object is empty or not, $\mathit{nextCalled \equiv lastRet \neq -1}$, which determines whether the \srct{remove} method of an \srct{Itr} object can be called (i.e., if \srct{next} has been called), $\mathit{mover \equiv size > cursor}$, which determines whether an \srct{Itr} has traversed all members of its corresponding \srct{Array} or not, and $\mathit{sync \equiv modCount = expectedModCount}$, which determines whether an \srct{Array} object has the same version as an \srct{Itr} object expects it (i.e., the \srct{Array} object has only been modified by the \srct{Itr} object).
Lastly, we specify integers as the domain of \srct{Array} objects.

\begin{figure}
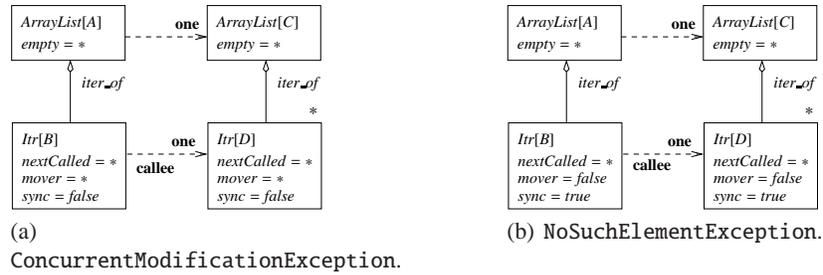
%
\centering

\subfigure[\srct{ConcurrentModificationException}.]{\input{figs/rolemap5.pstex_t}
\label{fig:rolemaps5}}
\hspace{7em}
\subfigure[\srct{NoSuchElementException}.]{\input{figs/rolemap6.pstex_t}
\label{fig:rolemaps6}}

\caption{Two exceptions for the \srct{next} method of \srct{ArrayList}.}
\label{fig:nextexcept}
\end{figure}

\begin{figure}
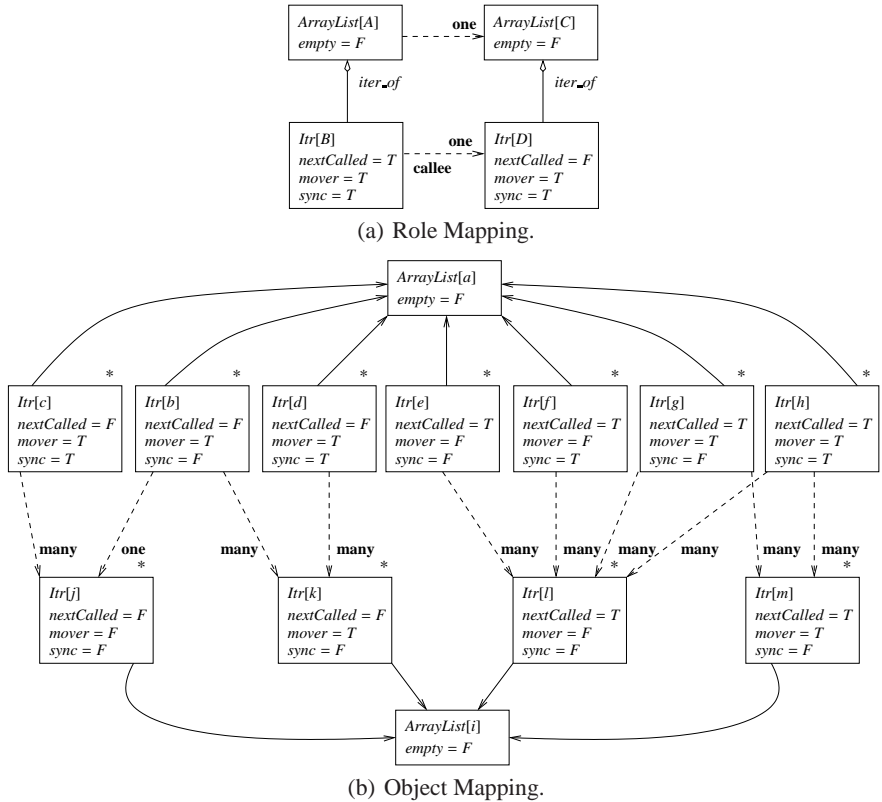
%
\centering

\subfigure[Role Mapping.]{\input{figs/rolemaps7.pstex_t}
\label{fig:rolemaps7}}
\hspace{2em}
\subfigure[Object Mapping.]{\input{figs/objmaps7.pstex_t}
\label{fig:objectmaps7}}

\caption{One of the three rules for \srct{remove} method of \srct{ArrayList}. ``$T$'' and ``$F$'' represent $true$ and $false$, respectively. For the sake of clarity the arrows representing reference predicates are not labelled with $\mathit{iter\_of}$.}
\label{fig:rule5}
\end{figure}

Our tool can compute 15 rules that cover all possible behaviour of \srct{ArrayList}. 
It once missed computing the rule for \srct{next} when called on an iterator whose all predicates are true and remain true after the method call.
Figure \ref{fig:nextexcept} shows two exceptions rules that our tool computes for the \srct{next} method.
Figure \ref{fig:rolemaps5} shows the case when the \srct{next} method raises \srct{ConcurrentModificationException} because an \srct{Itr} callee object is not \emph{sync}.
Figure \ref{fig:rolemaps6} specifies when the \srct{NoSuchElementException} exception is raised.
Figure \ref{fig:rule5} specifies one of the three rules that our tool computes for the \srct{remove} method in one of our experiments.
This rule is interesting in that it demonstrates that the object mapping of a rule can be non-deterministic: e.g., a mover, sync iterator object can either become non-mover or stay mover, in both cases it becomes non-sync.
This rule could have been more general, however. 
First, in the source nested object graph, the object with $\mathit{nextCalled = false}$, $\mathit{mover = false}$, and $\mathit{sync = false}$ is missing.
Second, the object mapping from $b$ to $j$ could have had multiplicity ``many''.
And lastly, there could have been an object mapping from $d$ to $l$ with multiplicity ``many'' denoting that some of the mover, sync objects whose $\mathit{nextCalled}$ is false become non-movers.

\paragraph{HashSet.} 
The input for computing the DPI of \srct{HashSet} is somewhat different from \srct{ArrayList}'s. 
The \srct{HashSet} class has a \srct{map} field, which is a \srct{HashMap}.
Most of the semantics of \srct{HashSet} is implemented via \srct{map} and its methods.
In particular, an iterator for a \srct{HashSet} object, is an inner \srct{HashIterator} object of its \srct{map} object.
Furthermore, two of the input predicates are also defined differently: $\mathit{mover \equiv next \neq null}$ and $\mathit{nextCalled \equiv current \neq null}$.
Using these input information, our tool computed 15 rules: the same number of rules as for \srct{ArrayList}. 
Upon a closer examination, we noticed two differences between the two rules of the two packages.
First, while in \srct{ArrayList} rules for invoking the \srct{add} method on an \srct{Array} object causes \emph{all} iterators that point to it to become unsync (which made sense because the other iterators should become invalid), in the case of the rules for \srct{HashSet} some sync iterators would become unsync, while the others would remain sync.
The reason turned out to be that adding a duplicate element to a \srct{HashSet} object does not change the \srct{modCount} attribute of the object, and thus a sync iterator would remain sync. 
Our tool, however, had merged rules for adding duplicate elements with rules for adding new elements, leading to a mix of sync and unsync iterator objects as a result.
Method \srct{add}, however returns a false value if it receives a duplicate value.
We adjusted our input to the tool so that \srct{add} rules with distinct return values are distinguished. 
By default, our tool abstracts away from the return value of modifier methods, because they are in general not useful to distinguish genuinely different rules; e.g., the return value of \srct{next} method of an iterator returns an object, which can have no role in distinguishing between genuinely different rules.
With this new input, our tool computed a set of 16 rules, which distinguish between the case when a new element is added to a non-empty set and the case when a duplicate element is added to a non-empty set.

The second difference is that the $\mathit{mover}$ predicate of an \srct{Iterator} object of a \srct{HashSet} only correctly denotes whether it has traversed all elements of its corresponding \srct{HashSet} or not if its $\mathit{sync}$ predicate is true.
This is essentially because unlike an \srct{ArrayList} object, whose iterator objects maintain an index of the underlying array of the \srct{ArrayList} object, the iterators of a \srct{HashSet} need to traverse the underlying hash table of the \srct{HashSet}, which is not contiguously populated.

Lastly, from Table \ref{tab:results}, it is clear that computing the DPI of \srct{HashSet} takes significantly longer than computing the DPI of \srct{ArrayList}.
This difference can be partly justified by the fact that \srct{ArrayList} implements \srct{RandomAccess}, which provides constant-time access, while \srct{HashSet} does not.
Another slowing factor is the way the reference predicates are computed for the two packages.
For \srct{ArrayList}, we only need to check which \srct{Array} object an \srct{Itr} object resides in.
For \srct{HashSet}, we can check which \srct{HashMap} a \srct{HashIterator} resides in, but then we need to check which \srct{HashSet} object wraps that \srct{HashMap} object.
To find that \srct{HashSet} object we need to check all objects of the snapshot and perform reflection on their type and their \srct{map} fields.

\section{Conclusion}\label{sect:conclusion}
We have introduced the notion of dynamic package interface (DPI) that provides a succinct way to 
describe valid usage patterns for a package. 
The DPI of a package is a set of rules, each of which specifies the effect of a method call over a 
general configuration of a set of objects.
We have developed a dynamic tool that computes an approximation of the DPI of a Java package automatically, 
given a set of abstraction predicates.
The rules of such a DPI generalize the usual examples used in the documentation of the 
Java package and can be traced to problems discussed in online forums.

A DPI captures both the \emph{inter}-object aspects 
of the dynamic behaviour of the classes of a package, as well as the \emph{intra}-object aspects of individual classes of the package, 
relative to a set of scalar and reference predicates, even when unboundedly many objects interact.\footnote{
We use the terms ``inter-object'' and ``intra-object'' in a similar sense as in OO design~\cite{Damm01:LCS}.}
In contrast, previous dynamic techniques primarily focus on either deriving intra-object specifications for one object
or deriving finite state machines that capture the interaction pattern of a 
finite number of objects \cite{Ghezzi09:Synthesizing,Dallmeier10:Generating,Pradel12:Static,Pradel09:Automatic,Henkel07:Discovering,Nguyen09:Graph}.

Lastly, our work focuses on the analysis of the classes of packages that are non-recursive; it abstracts away from their underlying recursive data structures, which are often only accessible internally or privately via the public classes of the package. 
Our analysis can be considered as orthogonal to the techniques, both dynamic and static, that deal with recursive data structures.

\bibliographystyle{splncs03}

\bibliography{main}

\end{document}